\documentclass[12pt,preprint]{aastex}
\begin{document}

\shorttitle{IRAC observations of MASH planetary nebulae} %shortened to fit ApJ entry box
\shortauthors{Cohen et al.}

\title{{\it Spitzer} IRAC observations of newly-discovered planetary nebulae 
from the Macquarie-AAO-Strasbourg H$\alpha$ Planetary Nebula Project}

\author{Martin Cohen\altaffilmark{1}, Quentin A. Parker\altaffilmark{2,3}, Anne J. 
Green\altaffilmark{4}, Tara Murphy\altaffilmark{4,5},\\ 
Brent Miszalski\altaffilmark{2,6}, David J. Frew\altaffilmark{2,7}, Marilyn R. 
Meade\altaffilmark{8},\\
Brian Babler\altaffilmark{8}, R\'emy Indebetouw\altaffilmark{9}, Barbara A. 
Whitney\altaffilmark{10},\\
Christer Watson\altaffilmark{11}, Edward B. Churchwell\altaffilmark{8},
and Douglas F. Watson\altaffilmark{8}}

\altaffiltext{1}{Radio Astronomy Laboratory, University of California,\\ 
Berkeley, CA 94720; {\bf mcohen@astro.berkeley.edu}}
\altaffiltext{2}{Department of Physics, Macquarie University, Sydney,\\ 
NSW 2109, Australia}
\altaffiltext{3}{Anglo-Australian Observatory, PO Box 296, Epping,\\ 
NSW 2121, Australia}
\altaffiltext{4}{School of Physics, University of Sydney, NSW 2006,\\ 
Australia}
\altaffiltext{5}{School of Information Technologies, University of\\
Sydney, NSW, 2006, Australia}
\altaffiltext{6}{Observatoire Astronomique, Universit\'e Louis Pasteur,\\
67000, Strasbourg, France}
\altaffiltext{7}{Perth Observatory, Walnut Road, Bickley WA 6076,\\ 
Australia}
\altaffiltext{8}{Dept. of Astronomy, University of Wisconsin,\\ 
Madison, WI 53706}
\altaffiltext{9}{Astronomy Dept. University of Virginia,\\ 
Charlottesville, VA 22904}
\altaffiltext{10}{Space Science Institute, Boulder, CO 80303}
\altaffiltext{11}{Dept. of Physics, Manchester College, North\\  
Manchester, IN 46962}

%\date{ Accepted . Received ; in original form        }

%\renewcommand{\baselinestretch}{1.6}

\begin{abstract}
We compare H$\alpha$, radio continuum, and {\it Spitzer} Space Telescope (SST) images
of 58 planetary nebulae (PNe) recently discovered by the Macquarie-AAO-Strasbourg H$\alpha$   
PN Project (MASH) of the SuperCOSMOS H$\alpha$ Survey. Using InfraRed Array Camera (IRAC) data
we define the IR colors of PNe and demonstrate good isolation between these colors and those
of many other types of astronomical object.  The only substantive contamination of PNe in
the color-color plane we illustrate is due to YSOs.  However, this ambiguity is readily
resolved by the unique optical characteristics of PNe and their environs.
We also examine the relationships between optical and MIR morphologies from 3.6 to 8.0\,$\mu$m,
and explore the ratio of mid-infrared (MIR) to radio nebular fluxes, which is a valuable
discriminant between thermal and nonthermal emission.  MASH emphasizes late
evolutionary stages of PNe compared with previous catalogs, enabling study
of the changes in MIR and radio flux that attend the aging process.
Spatially integrated MIR energy distributions were constructed for all MASH PNe observed by
the GLIMPSE Legacy Project, using the H$\alpha$ morphologies to establish the
dimensions for the calculations of the Midcourse Space Experiment (MSX),
IRAC, and radio continuum (from the Molonglo Observatory Synthesis
Telescope and the Very Large Array) flux densities. The ratio of
IRAC\,8.0-$\mu$m to MSX 8.3-$\mu$m flux densities provides a measure of the
absolute diffuse calibration of IRAC at 8.0\,$\mu$m.  We independently confirm the
aperture correction factor to be applied to IRAC at 8.0\,$\mu$m to align it with the diffuse
calibration of MSX.  The result is in accord with the recommendations of the {\it Spitzer}
Science Center, and with our results from a parallel study of H{\sc ii} regions
in the MIR and radio.  However, these PNe probe the diffuse calibration of IRAC on a
spatial scale of 9$^{\prime\prime} - 77^{\prime\prime}$, as opposed to the
many arcmin scale from the study of H{\sc ii} regions.
\end{abstract}


\section{Introduction}
The recent availability of the SuperCOSMOS AAO/UKST H$\alpha$ survey of the Southern
Galactic Plane; SHS (Parker et al. 2005) has led to a substantial 60\% increase in
the numbers of Galactic planetary nebulae (PNe) identified. This is largely
due to the excellent combination of resolution ($\sim1$~arcsecond), sensitivity
($\sim5$~Rayleigh) and areal coverage (4000~sq.degrees) offered by this
powerful online survey\footnote{http://www-wfau.roe.ac.uk/sss/halpha/}.
These 905 new PNe have recently been published as the
Macquarie/AAO/Strasbourg H$\alpha$ PN project (MASH) by Parker et al. (2006).
In this paper we describe a multiwavelength
examination of a subset of these objects for which {\it Spitzer} Space Telescope
(Werner et al. 2004: hereafter SST) observations are available between 3.6 and 8.0\,$\mu$m
from the GLIMPSE survey of portions of the Galactic Plane (Benjamin et al. 2003;
Churchwell et al. 2004).  We compare multiwavelength images of PNe in H$\alpha$,
the 8.3-$\mu$m band of the Midcourse Space Experiment (MSX: Price et al. 2001), the
four InfraRed Array Camera (IRAC: Fazio et al. 2004) bands, and radio continuum from
either the Molonglo Galactic Plane Survey in the south (MGPS2: Green 2002) or the NVSS in the 
north (Condon et al. 1998).

MASH PNe lie within Galactic latitudes of $|{\it b}|$ $10^\circ-13^\circ$.  MSX imaged the entire
Galactic Plane within $\pm$5$^\circ$ while GLIMPSE is confined to $\pm$1$^\circ$.  For
the present multiwavelength study of PNe we investigated both regions of
the GLIMPSE survey, covering the longitudes from 10$^\circ$ to 65$^\circ$ and 295$^\circ$ to 350$^\circ$.
As part of the final culling of MASH candidates prior to publication, various multiwavelength comparisons 
were made between MASH PN candidates and other extant data from MSX, 2MASS and, for the area of 
overlap, GLIMPSE. There were 85 PNe candidates in the MASH/GLIMPSE overlap zone to begin with.
Mid-infrared (MIR) data are valuable to study the properties of these objects as well as
to remove misidentifications from the MASH (Cohen \& Parker 2003). 
Parker et al. (2006) discuss in detail how non-PN contaminants were eliminated
from the MASH catalog.  An environment with the indicators of active star-formation or heavy dust
obscuration, or an optical morphology showing multiple patches of nebulosity, possibly linked,
suffice to remove many H{\sc ii} regions.  Any object that lacks H$\alpha$ emission will appear
the same in the narrowband H$\alpha$ and accompanying broadband red continuum exposures,
precluding reflection nebulae and normal galaxies.  Compact H{\sc ii} regions were eliminated
by optical spectroscopy based on the weakness of [N{\sc ii}](6548+6584) relative to H$\alpha$ emission
(hereafter [N{\sc ii}]/H$\alpha$).  Additional emission-line objects that are not PNe were rejected 
by optical spectroscopy if they lacked the characteristic PN lines (e.g. [O{\sc iii}] in the blue; 
[O{\sc i}], [N{\sc ii}], and [S{\sc ii}] in the red) or showed a continuum.
MIR morphology is often sufficient to discriminate between PNe and H{\sc ii} regions.  

MASH PNe are generally distinguishable from previously known nebulae by their larger size and faintness.
This distinction translates to large proportions of highly evolved objects of large apparent
size and low surface brightness that have almost dissolved into the interstellar medium (ISM),
and of compact faint PNe.  This complementarity between the MASH and old PN catalogs is 
important because only a small fraction of the estimated $\sim$30,000 Galactic PNe (e.g. Frew \&
Parker 2005) is known.
To understand PNe it is vital to identify more of this population and to have access to a broad
representation of different evolutionary stages.  This helps to gauge the relationship
between morphology and age and to quantify the gradual chemical enrichment of the Galaxy.

Careful consideration of MASH optical spectra, and of MSX and {\it Spitzer} imagery of MASH PN candidates
resulted in the rejection of a total of 27 candidates: 4 objects were found to correspond to portions of
known supernova remnants; 4 are inconclusive identifications following consideration of the available optical
and IR data; one is a symbiotic star; 2 are ``likely" PNe; and the remaining 16 are probably H{\sc ii} regions and
were anyway never classified as more than ``possible" PNe. Fifty-eight nebulae survived this cull, composed of 28 true,
16 likely, and 14 possible PNe  (from the classification categories of Parker et al. 2006).
Objects inconclusive in nature have been placed in the MASH Miscellaneous
Emission Nebulae (MEN) catalog until their true character can be revealed.
Some of these may later turn out to be PNe and will return to MASH.

In this paper we present the sample of PNe and summarize their multiwavelength characteristics (\S2);
describe the tools we use to confirm true MIR counterparts of the 58 objects (\S3);
compare the ratio of IRAC 8.0-$\mu$m to MSX 8.3-$\mu$m fluxes to investigate further the
accuracy of IRAC's absolute diffuse calibration on smaller spatial scales than probed by
Cohen et al. (2006) in their study of H{\sc ii} regions (\S4); compare the IRAC colors of the 
MASH PNe with those found by Hora et al. (2004) to provide a diagnostic color-color plot which
distinguishes PNe from other sources in the GLIMPSE data (\S5);
examine the ratio of spatially-integrated MIR/radio flux density for PNe (\S6); and 
compare MIR and optical morphologies of PNe (\S7).  \S8 gives our conclusions.

\section{The sample of PNe}
\subsection{Removing non-PNe from the MASH}
\begin{figure}
\includegraphics[scale=0.41]{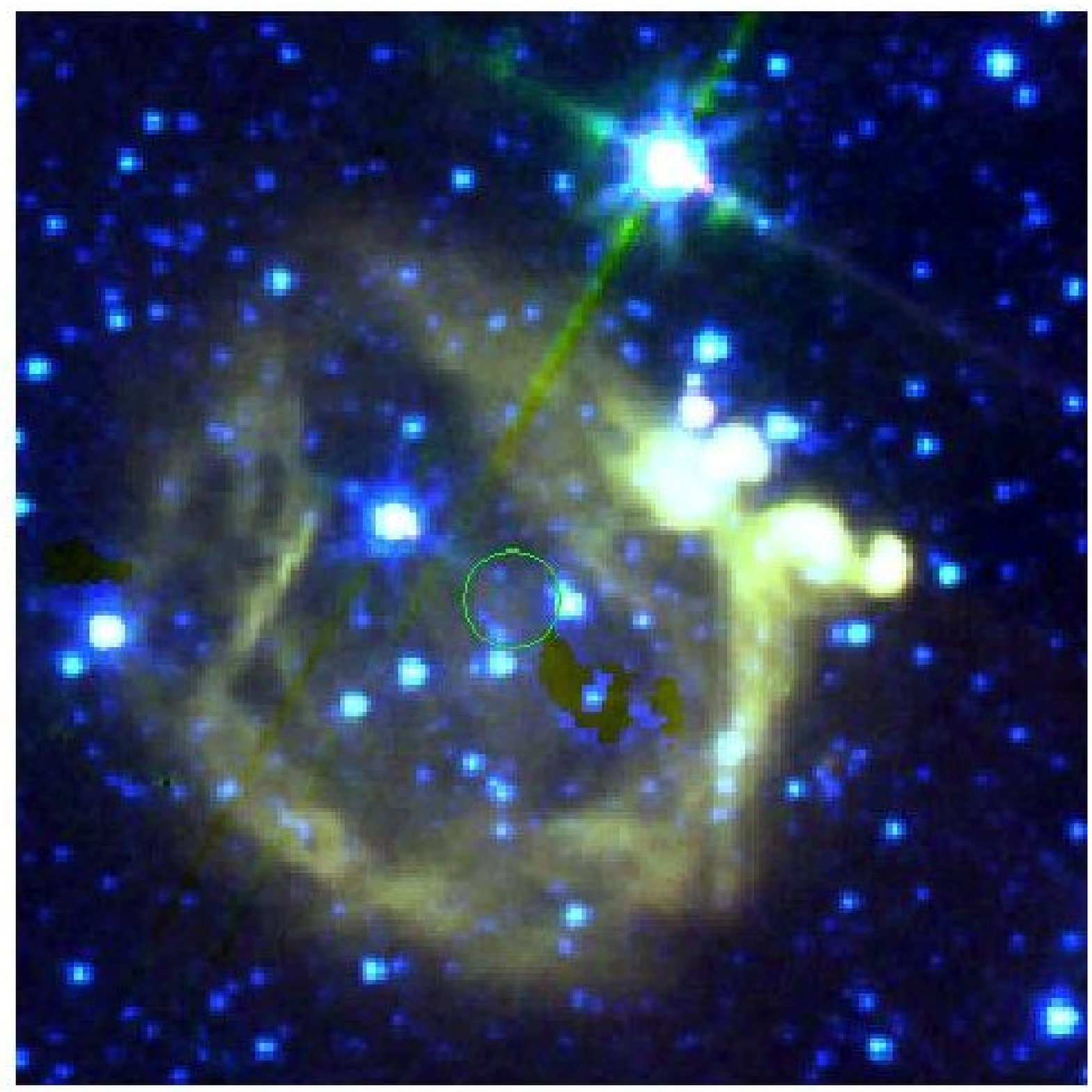} %f32139m031_234.eps}
\includegraphics[scale=0.40]{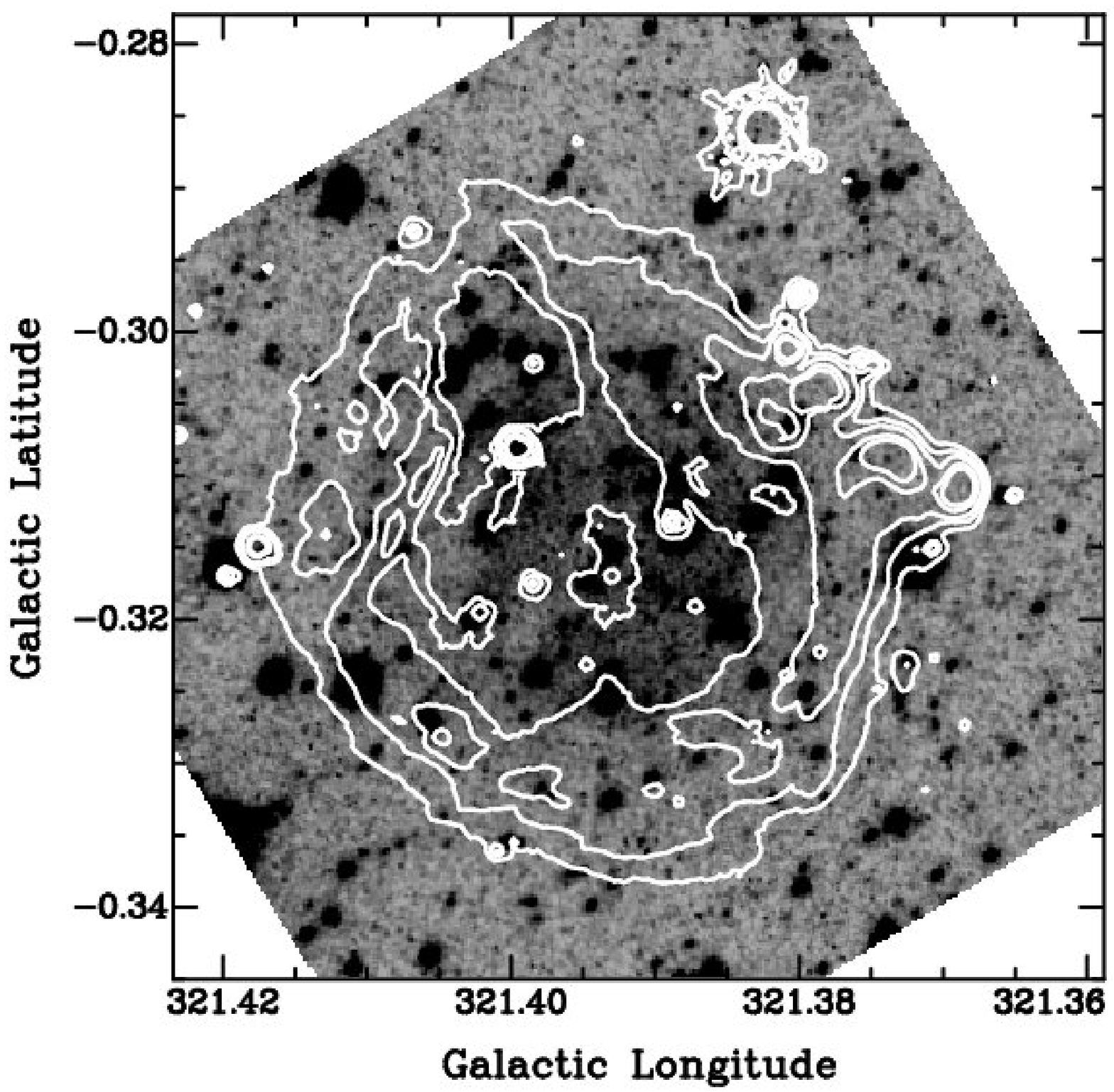} %1517-5751ha_8dkr.eps}
\caption{Left: MIR false color morphology of the IRAC counterpart to PHR1517$-$5751 %preprint
%\figcaption{Top: MIR false color morphology of the IRAC counterpart to PHR1517$-$5751 %2column
(blue, green, and red represent emission at 4.5, 5.8, and 8.0\,$\mu$m, respectively).
Image is 4$\times$4\arcmin. The object consists of an almost complete 
ring which has the appearance of an H{\sc ii} region rather than a PN.  
Center of the small circle (12$^{\prime\prime}$ radius) is the MASH  
%center of the nebula. Bottom: H$\alpha$ image of the same PN covering the same area %2column
centroid of the nebula. Right: H$\alpha$ image of the same PN covering the same area %preprint
as shown in the IRAC image, overlaid by white contours of 8.0-$\mu$m emission at 
levels of 50, 70, 100, 200, 300\,MJy\,sr$^{-1}$. \label{nonpn}}  %\label{1517ha}
\end{figure}

\begin{figure}
\includegraphics[scale=0.60,angle=-90]{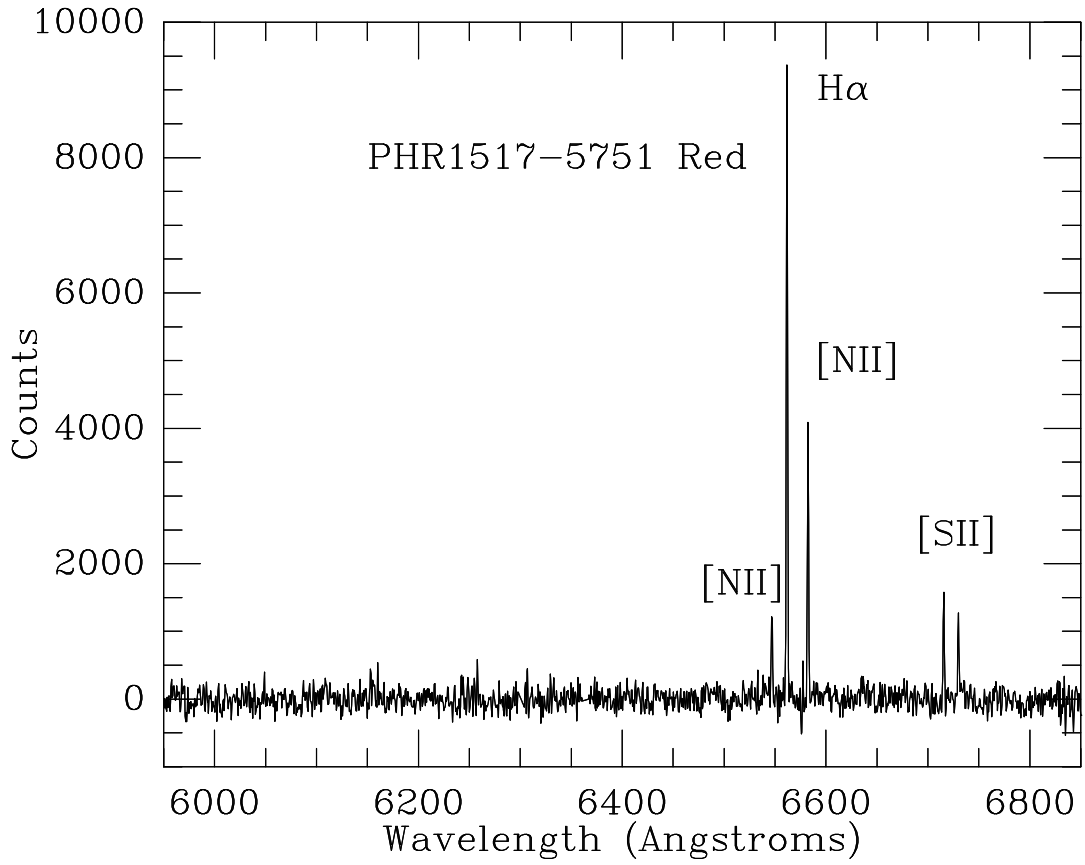} %1517r.eps}
\caption{Red spectrum of PHR1517$-$5751 showing the weakness of the [N{\sc ii}] lines
relative to H$\alpha$ that define this object as an H{\sc ii} region. \label{1517rspec}}
\end{figure}

Fig.~\ref{nonpn} compares the H$\alpha$ appearance with the IRAC morphology of a PN candidate that has now 
been reclassified as an H{\sc ii} region. % (Fig.~\ref{1517ha}.  
It was initially considered a PN candidate because of its plausible optical morphology and its
reasonable optical red spectrum (Fig.~\ref{1517rspec}, with [N{\sc ii}]/H$\alpha$) of 0.60, close to
the cut-off value for H{\sc ii} regions, of 0.70 (Kennicutt et al. 2000).  High extinction 
precluded blue spectroscopy that might have revealed the absence of PN lines.  However, the relative locations of H$\alpha$ 
and 8-$\mu$m emission (see Fig.~\ref{nonpn}), and what appear to be possible secondary star formation
regions (several compact H{\sc ii} regions on the north-western rim) are much more consistent with
an H{\sc ii} region.  

\begin{figure} 
\includegraphics[scale=0.44]{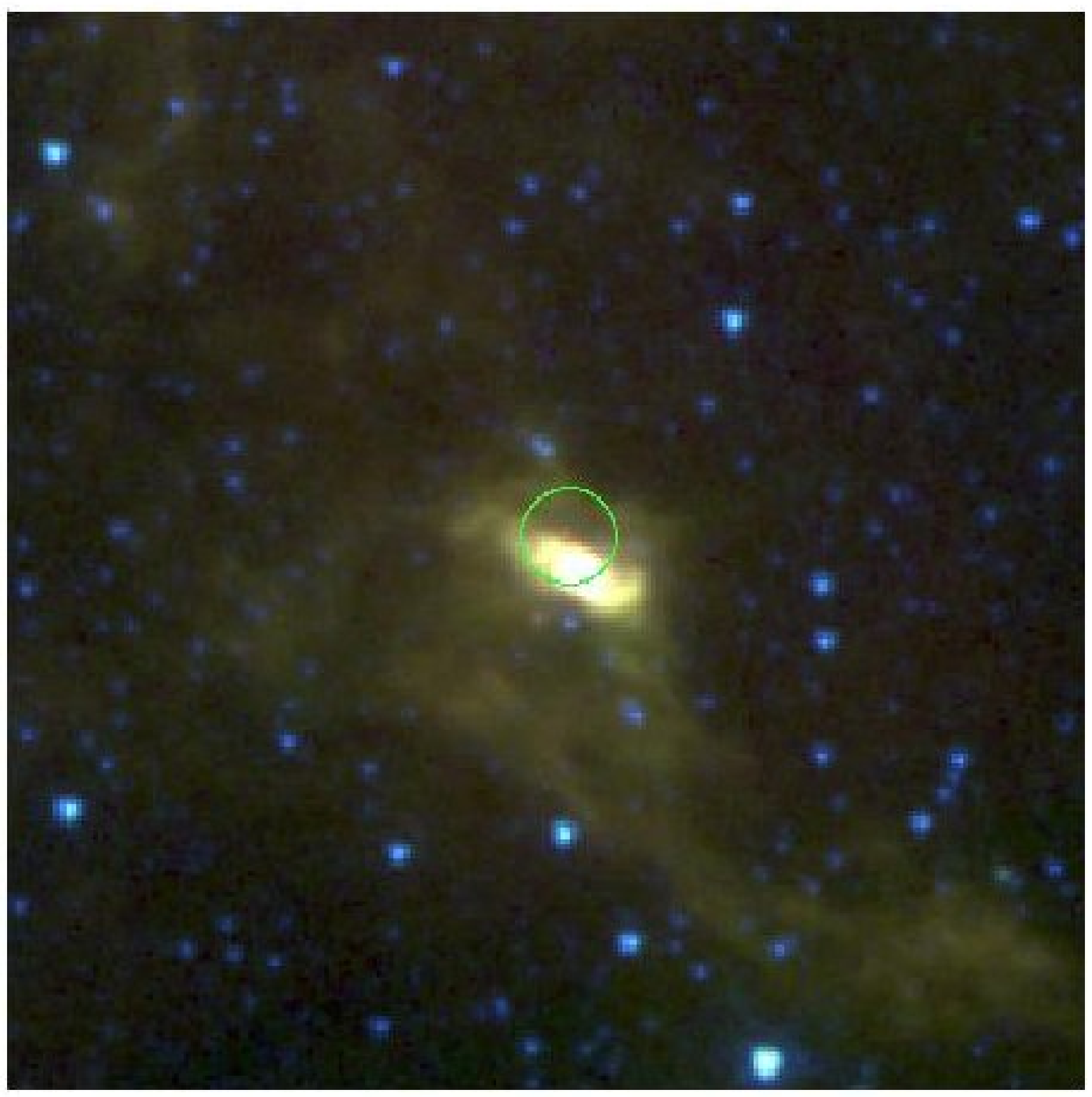} %n30311m096_234.eps}          
\includegraphics[scale=0.40]{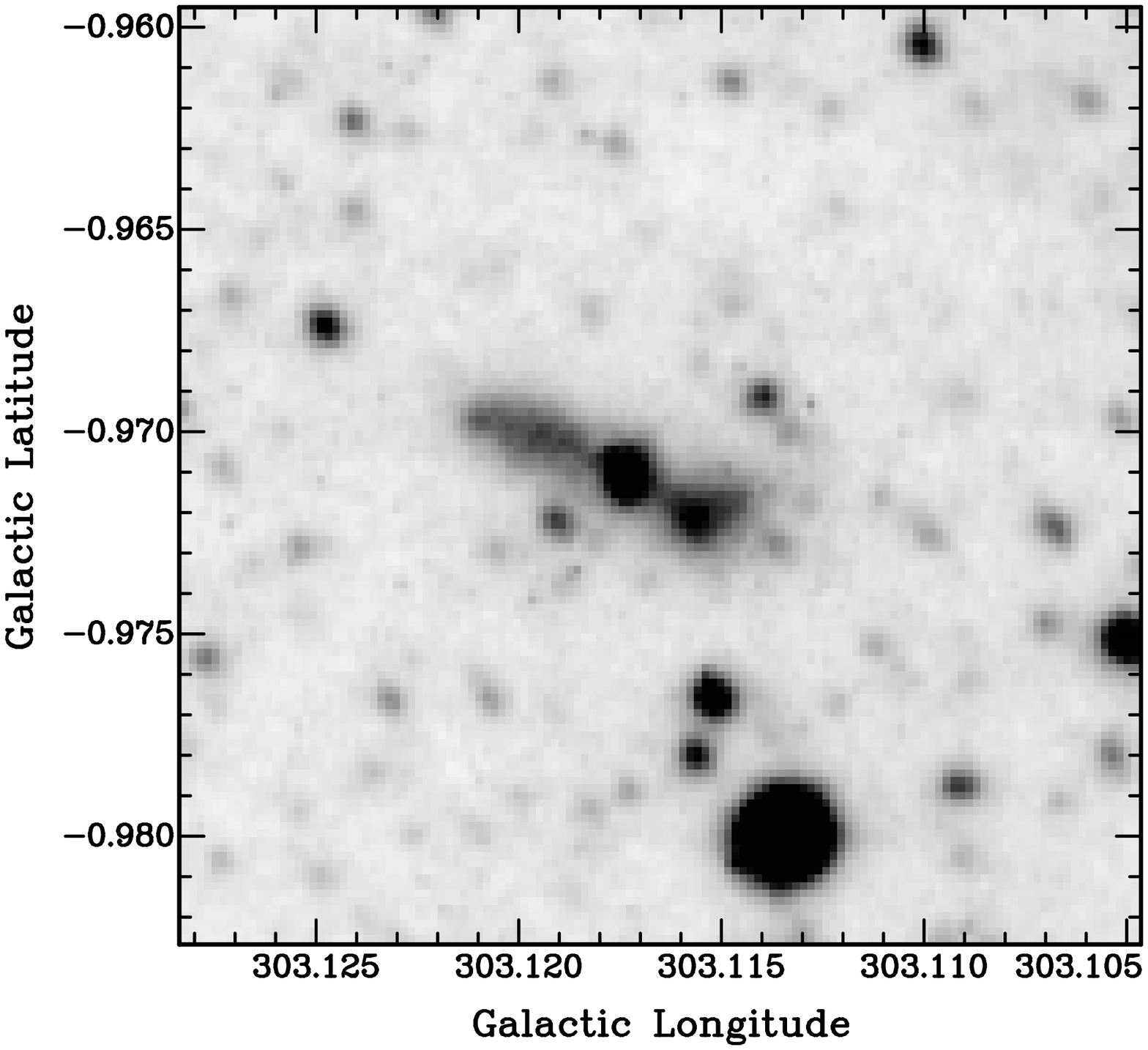} %1253-6350ha.eps}     
\caption{Left: MIR false color image (as in Fig.~\ref{nonpn}) of the %preprint
%\caption{Top: MIR false color image (as in Fig.~\ref{nonpn}) of the %2column
bipolar counterpart to PHR1253$-$6350.  Image is 4$\times$4\arcmin.
Note the several bright MIR filaments that extend for several arcmin. Circle and image size 
as for Fig.~\ref{nonpn}. Right: H$\alpha$ blow-up showing nebular detail in the form %preprint
%as for Fig.~\ref{nonpn}. Bottom: H$\alpha$ blow-up showing nebular detail in the form %2colum
of two asymmetric bipolar lobes flanking the image of the symbiotic star.\label{wingnut}} 
\end{figure}

\begin{figure}
\includegraphics[scale=0.50,angle=-90]{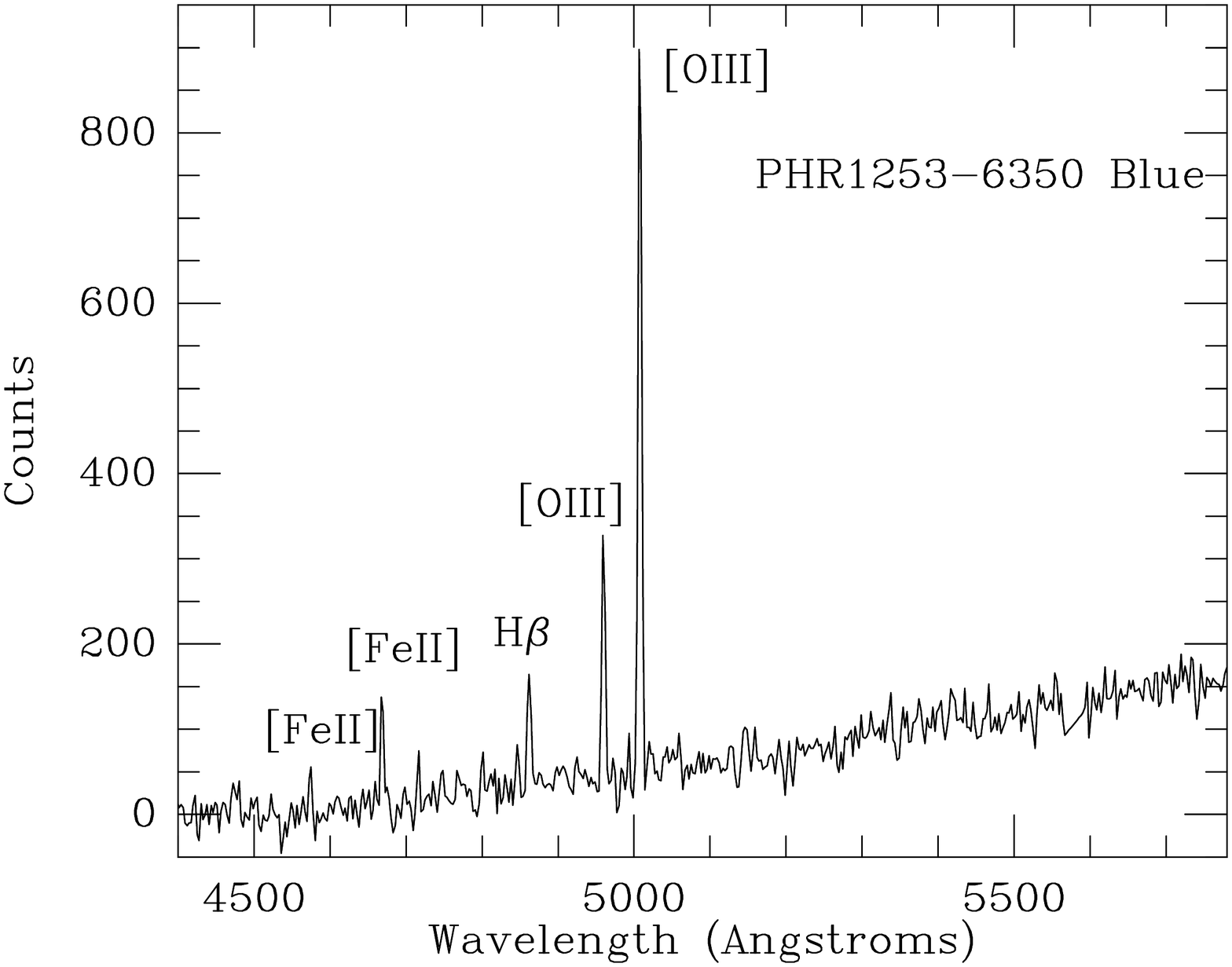} %1253b.eps}
\includegraphics[scale=0.50,angle=-90]{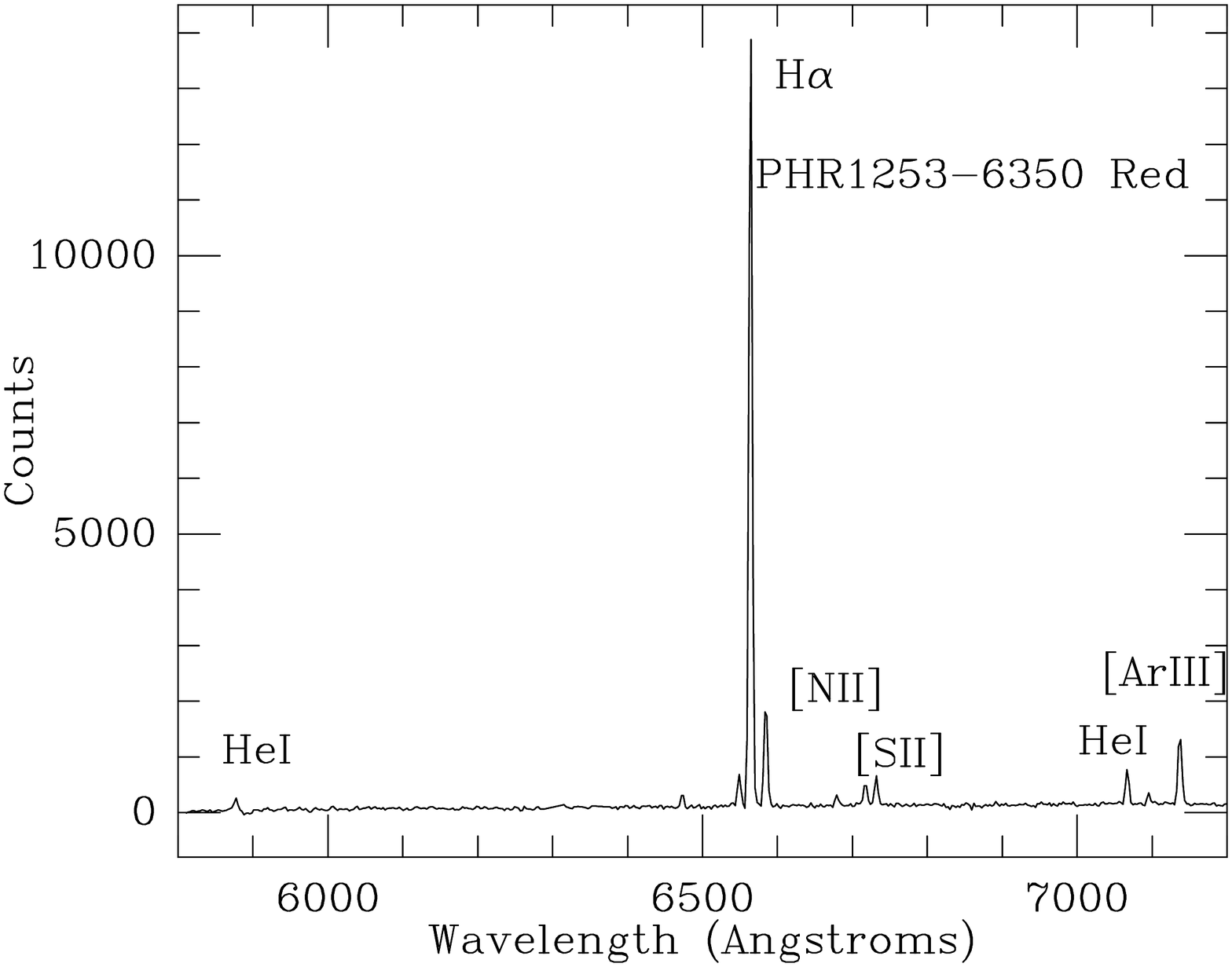} %1253r.eps}
\caption{Optical spectra of PHR1253$-$6350.  Top: blue spectrum showing
[Fe{\sc ii}] lines at 4575 and 4665\AA, together with nebular lines of H$\beta$
and [O{\sc iii}] on an obvious continuum.  Bottom: red spectrum showing a faint red continuum 
with faint [Fe{\sc ii}] lines at 6474 and 7094\AA;
He{\sc i} lines (5876, 6678, 7065\AA); H$\alpha$; the red nebular lines of [N{\sc ii}],
[S{\sc ii}], and [Ar{\sc iii}] at 7136\AA. \label{1253spec}}
\end{figure}

Further examples of more subtle contaminants in the MASH are found among highly collimated bipolar nebulae
that contain dusty equatorial disks producing thermal emission from warm dust with
temperatures typically in the range 250-400~K.  One such interesting object is PHR1253$-$6350, 
previously identified as a possible highly collimated bipolar PN with a bright central star (CS). 
Fig.~\ref{wingnut} presents both MIR and H$\alpha$ morphologies.  (The MIR image is a 3-band
false color image as described in \S3.2.)  The nebular spectrum (Fig.~\ref{1253spec})
is PN-like with [O{\sc iii}]/H$\beta$$\approx$10, [N{\sc ii}]/H$\alpha$$\approx$0.1, but
with an obvious blue, and a weaker red, continuum.  The CS is not blue but is bright in 
the near-infrared (NIR) with 2MASS photometry yielding $J-H$$\sim$1.43 and $H-K$$\sim$1.41.  Combining
these with a magnitude at 3.6\,$\mu$m of 7.35, and $K-$[3.6] of 2.87 indicates that the
stellar photosphere is overwhelmed by thermal emission from $\sim$700\,K dust (see 
Allen \& Glass 1974, their Fig.~2).   
The optical outflows coincide exactly with a strong elongated emission ridge of a much more
extensive underlying MIR structure which extends in filamentary form for  
several arcminutes from NE-SW. These structures are clearly related and certainly atypical of PNe,
as are the rising continua.
In its H$\alpha$ structure and the shape of its MIR core, this nebula is strongly redolent of the 
highly collimated, high-density nebulae discussed  by Corradi (1995).  He linked these systems to 
interacting binaries containing a symbiotic star.  Among the usual nebular lines, and those from H and He{\sc i}, 
one sees four [Fe{\sc ii}] lines that are associated with symbiotic spectra.  Hence a more likely 
identification would be a highly collimated bipolar outflow, perhaps from a symbiotic nucleus.
Characterizing PHR1253$-$6350 as a bipolar symbiotic outflow does not explain
the relationship between this object and its environment.  Does the geometry of the extended MIR
emission indicate the passage of this star through the ISM, represented
by the filaments?  Or was the object born within the material traced by the halo of MIR emission that
surrounds it and that is excited by the bright bipolar nebula inside?
This intriguing object is the subject of a separate paper (Parker, Cohen \& Frew, in preparation).
Although we regard it as a non-PN, the influence of environment on the evolution of PNe is still a
relevant issue, as shown by the work of Villaver et al. (2006) on the potential contribution
of the ISM to PN morphology.

\begin{figure}
\includegraphics[scale=0.60,angle=-90]{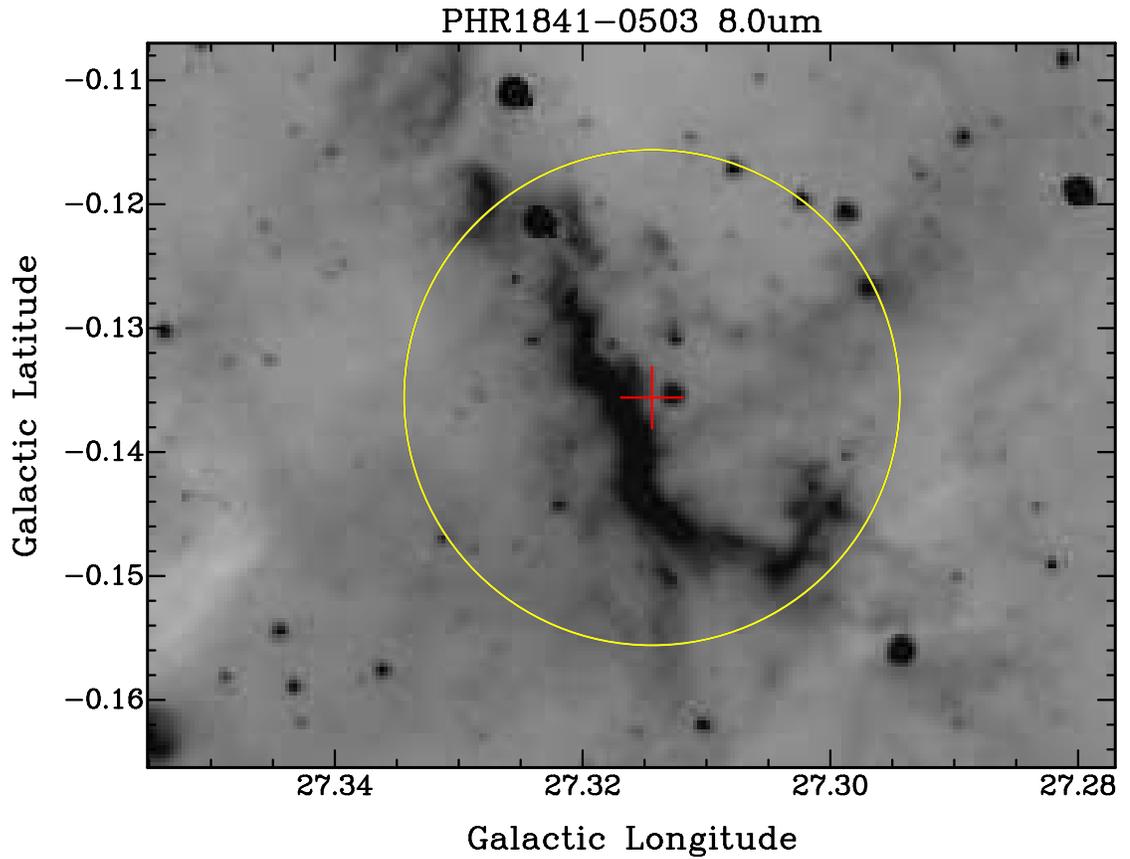} %PHR1841-0503.eps}
\caption{MIR image at 8.0\,$\mu$m of a probable sinuous PDR showing the location of 
the PN candidate, PHR1841$-$0503, just off the ridge.  The cross has
9$^{\prime\prime}$ arms; the large circle has a 72$^{\prime\prime}$ radius). \label{ridge}}
\end{figure}

One caveat relates to the constraint imposed by the GLIMPSE survey parameters on MASH PNe
in this paper.  The narrow latitude range, around the Galactic Equator, of GLIMPSE should result in
relatively few PNe because of the heavy extinction.  The MASH survey has found
new PNe in the plane because of the excellent sensitivity of the AAO/UKST H$\alpha$ survey, where the 
longer central wavelength of the interference filter compared with more traditional PN
search techniques based on [O{\sc iii}] detection, is less affected by extinction.  Part of this
success also stems from  the evident patchiness of the optical 
obscuration.  However, only 6\% of the MASH catalogued PNe lie within one degree of zero latitude.  
In fact, one might well identify an H$\alpha$ emission region along an unusually
clear line-of-sight as an isolated PN when it is merely a less obscured part of a larger entity.  A good
example of this phenomenon that again emphasizes the value of MIR maps of the plane is shown by
Fig.~\ref{ridge}.  PHR1841$-$0503's location is identified by the cross in this 8.0-$\mu$m {\it Spitzer}
image.  One can now recognize that the candidate corresponds only to a local optical brightening along
an extended ridge of polycyclic aromatic hydrocarbon (PAH)
 emission.  If the small westerly offset from the center of the MIR ridge
were significant then one might argue for the ridge to represent a
photodissociation region (PDR) created by ionizing radiation from some star(s) to the west of the
sinuous bright 8.0-$\mu$m filament.

\subsection{The available data}
Table~\ref{pnelist} contains the following information: Column (1) -- source name; Column (2) -- status of the PN
as true (T), likely (L), or possible (P); Columns (3-4) -- Galactic coordinates in degrees;
Columns (5-6) -- J2000 Equatorial coordinates with units shown; Column (7) --  optical dimensions in arcsec;
Column (8) -- optical morphology code, described in \S2.4 and in Parker et al. (2006); Column (9) -- shows
the presence (``y'') or absence (``n'') of a clear MIR false-color counterpart (from IRAC) to the PN; Column
(10) -- indicates if an IRAC counterpart (in any or all of the 4 bands) is linked morphologically to the
H$\alpha$ nebular image; Columns (11-13) --  list any  MSX 8.3-$\mu$m or radio continuum counterparts,
with the presence (``y'') or absence (``n'') of detections listed and ``..." indicating that no image from 
MGPS2 or NVSS was available; Column (14) -- notes if the likely central star of the PN is detected, with
``o'' implying an optically identified CS candidate in an appropriate position
(for three PNe that candidate is optically blue and is footnoted), ``i'' signifying a CS candidate 
found in 2MASS images, ``m" indicating a possible stellar candidate in at least one IRAC band, and ``N'' denoting 
nebular emission seen in 2MASS.  
One of the MASH PNe, PHR1815-1457, designated as a likely PN, lies so close to the upper limit of latitude in
the GLIMPSE coverage that it was unobserved at 3.6 and 5.8\,$\mu$m, although images are
available in the other band pair (4.5 and 8.0\,$\mu$m) because of the different fields
of view observed through IRAC's two dichroic beamsplitters.

When looking for the MIR counterpart morphologically, one may see a resolved source exactly matching
the ionized gas as traced by the H$\alpha$ image.  If this occurred in the two short IRAC wavelengths
(3.6 and 4.5\,$\mu$m) it would suggest that recombination lines dominated the emission in these two
filters (e.g. Pf$\gamma$ at 3.3\,$\mu$m and Br$\alpha$ at 4.05\,$\mu$m, respectively).
Alternatively, if there is 5.8 and 8.0-$\mu$m emission that follows the form of the ionized gas but with 
a greater extent, then we would attribute the MIR emission in these two bands to radiation from
fluorescing PAHs in a PDR that wraps around the PN.  Only five nebulae are
detected by 2MASS as extended NIR objects: PHR1457$-$5812, PHR1831$-$0805, PHR1843$-$0325,
PHR1857+0207, and PHR1619$-$4914 (PM\,5, the only known Galactic [WN] central star of a
planetary nebula (Morgan, Parker \& Cohen 2003)).

The position of a potential CS might be close to the optical centroid (i.e. the coordinates provided
in Table~1) for a ring, an elliptical or a bipolar nebula.  For arcuate or partial ring nebulae the
relevant location would be the center of curvature of the arc.  Ten of the PNe, or 17\% of the sample,
show optical CS candidates; 12\% NIR; and 24\% MIR.  

\subsection{GLIMPSE residual images}
Unlike with the MSX images, we have worked from GLIMPSE ``residual images".  These are
3.1$^{\circ}$$\times2.4^\circ$ images with 1.2$^{\prime\prime}$ pixels
from which all GLIMPSE point sources have been removed.  The residual images
are ideal for enhancing the recognition of diffuse nebulosity in regions of
high point source density and enable a far more reliable photometry of
such emission.  The residual images use our adaptation of {\sc daophot ii}
(Stetson 2000) for all GLIMPSE sources detected down to 2$\sigma$, deeper than the
publicly released Catalog and Archive point source lists that are
required to meet higher reliability criteria by sources being detected multiple
times at a 5$\sigma$ level in one or more channels.  No measurements less
than 3$\sigma$ are ever listed in either the GLIMPSE Catalogs or Archives.
Thus there are faint 2$-$3$\sigma$ point sources
that are subtracted from the residual images but which are not listed in GLIMPSE enhanced products.
Residual images may contain sources not extracted by {\sc daophot}, such
as saturated sources and sources that peak beyond the non-linearity limit for each
band.  These objects were analysed individually and their integrated flux densities
were subtracted.

\subsection{Nebular morphology descriptors}
A detailed description of the morphological classifications applied to the 
sample and some of the subsequent analysis form the basis of a separate paper (Parker
et al., in preparation). However, after review of the existing categories, an
adaptation of the current Corradi \& Schwarz (1995) scheme was employed, which is based
on a system developed by Schwarz, Corradi \& Stanghellini (1993).  A basic ``ERBIAS''
classifier is used to indicate PNe which are Elliptical, Round, Bipolar, Irregular,
Asymmetric or quasi-Stellar (point source). We add a distinction between elliptical and
circular PNe, based on interest in their canonical Str\"omgren spheres (e.g. Soker 2002)
where an object is considered to be Round (circular) if the difference between estimated
major and minor axes is $<$5\%. In uncertain cases a dual classification might be applied
such as ``E/B?". We then add a sub-classifier ``amprs'' to indicate an asymmetry ``a'',   
multiple shells or external structure ``m", point-like structure ``p'', a well-defined ring
structure ``r'' or resolved internal structure ``s''.  Typically, only one ``ERBIAS''   
classifier is given, but several ``amprs'' sub-classifications, listed alphabetically, may
be applicable. Our sample of 58 PNe is comprised of elliptical nebulae (51\%),
bipolar (28\%), round (9\%),  asymmetric (7\%), and irregular nebulae (5\%).  The
fraction of bipolar objects in the entire MASH is 12.5\%.  Limiting our sample
to the very low latitude coverage of GLIMPSE has more than doubled the fraction of 
bipolar nebulae in our sample.  

The correlation between highly bipolar PNe and strongly enhanced He and N abundances
is now well-known (e.g. Corradi \& Schwarz 1995) but the linkage between chemistry
and morphology was first noted by Greig (1967, 1971).  Subsequently Peimbert (1978)
and Peimbert \& Serrano (1980) defined Type~I PNe in terms of threshold values of
He or N abundance.  Most Type~I PNe were found to be bipolar (Peimbert 1978;
Peimbert \& Torres-Peimbert 1983), while surveys of bipolar PNe (Corradi \& Schwarz 1995)
confirmed their chemical peculiarities and added enhanced Ne abundance to these.
As a class, Type~I PNe have larger than average diameters and expansion velocities 
(Corradi \& Schwarz 1995), hotter and more massive CS (Tylenda 1989), smaller 
scale heights (e.g. Stanghellini 2000), and deviate from the circular 
rotation of the Galaxy.  Their association with more massive progenitors than typical
PNe is widely acknowledged on theoretical grounds too (e.g. Becker \& Iben 1980; Kingsburgh
\& Barlow 1994).  This is consistent with the high proportion of new Type~I PNe found
by the MASH, with its low-latitude coverage.

\section{MIR counterparts of the MASH PNe}
\subsection{8.0-$\mu$m IRAC imagery}
\begin{figure}
\includegraphics[scale=.65,angle=0]{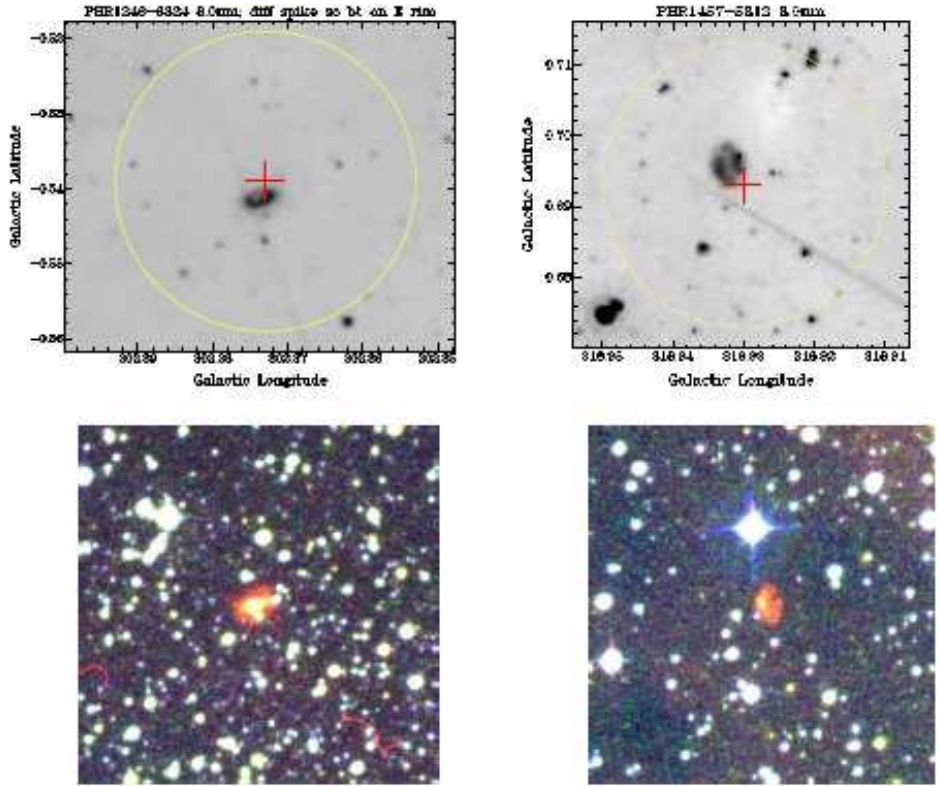} %pneha_8um.eps}
\caption{A pair of 8.0-$\mu$m IRAC images of the true PNe PHR1246$-$6324 and PHR1457$-$5812 (top),
compared with their respective H$\alpha$ images (below).  Dimensions of the small cross
and large circle as in Fig.~\ref{ridge}. \label{8umha}}
\end{figure}

To find the PAH emission commonly associated with carbon-rich PNe (e.g. Cohen et al. 1989)
we searched for MIR counterparts at 8\,$\mu$m. 
Hence, as part of the rigorous checking of all MASH candidate PNe prior to finalization of the
catalog, an initial check against available MIR data from MSX was made. This was later
supplemented at the lowest latitudes, by the new GLIMPSE data. As a consequence of this
combined effort a significant number of H{\sc ii} contaminants were culled from the MASH database
before its  publication. However, there are also a significant number of obviously 
spatially extended 8.0-$\mu$m  %(9 of 58) too few??!
counterparts of PNe.  Fig.~\ref{8umha} illustrates two such objects, both classified as true PNe:
PHR1246$-$6324 and PHR1457$-$5812.  Both were detected by MSX but no morphology could be discerned.
The figure compares MIR structures (upper images) with their corresponding H$\alpha$ images (lower images).  
PHR1246$-$6324 has a clear, tight, bipolar morphology while PHR1457$-$5812 is compact but
asymmetric.  Yet the similarities between the MIR and optical are clear.
The angular sizes of the MIR and H$\alpha$ images are in close agreement, with identical
position angles of the main axis of symmetry.  

\subsection{False-color {\it Spitzer} imagery} 
MIR spectra of PNe are not always dominated by PAH emission bands.  To eliminate the potential bias 
against any PN for which thermal emission from dust or ionic fine-structure lines might be the
major contributor to radiation in the IRAC bands we made a second comparison. False-color images
were produced using IRAC's 4.5, 5.8, and 8.0-$\mu$m bands shown in blue, green, and red, respectively.
This particular trio avoids the clutter due to the high stellar density in the 3.6-$\mu$m band.

The MSX data at 8.3 and 12.1\,$\mu$m also measure PAH emission, but 12.1-$\mu$m was generally not sensitive
enough to provide a confirming detection. By contrast, the IRAC results offer three sensitive bands
capable of sampling PAH emission (3.6, 5.8 and 8.0-$\mu$m). The 3.3-$\mu$m PAH band lies within IRAC's 3.6-$\mu$m band
but its strength is typically only 10\% of that of the 7.7-$\mu$m PAH band, while the 6.2-$\mu$m PAH feature
attains 56\% of the 7.7-$\mu$m band in PNe, reflection nebulae, and H{\sc ii} regions (Cohen et al. 1986;
their Table~5).  Therefore, 3.6-$\mu$m imagery is not a sensitive tracer of PAHs, 
and we looked for similar morphologies in the 5.8 and 8.0-$\mu$m bands to confirm the existence of PAHs, 
that are often the dominant spectral features in PNe. 
Furthermore, it is well-known that PNe also produce emission from lines of H$_2$
(ro-vibrational and pure rotational lines; e.g. Cox et al. 1998).  H$_2$ lines can contribute to emission 
in several IRAC bands.  They are chiefly seen at 8.0~$\mu$m 
(Hora et al. 2004) but can be responsible for over 90\% of the flux detected in IRAC's 3.6, 4.5, and 5.8-$\mu$m
bands (Hora et al. 2006).  This molecular emission in PNe can arise either in the warm molecular zone of a PDR
or in shocks.  In NGC\,6302 the molecular gas is photo-dissociated (Bernard-Salas \& Tielens 2005) while, in
the main ring of the Helix, 90\% of the H$_2$ emission is in shocks and only 10\% in PDRs (Hora et al. 2006).
Such PNe could appear red but might have different morphologies at 5.8 and 8.0\,$\mu$m, rather than identical
structures as with pure PAH emission.

\begin{figure}
\includegraphics[scale=0.75]{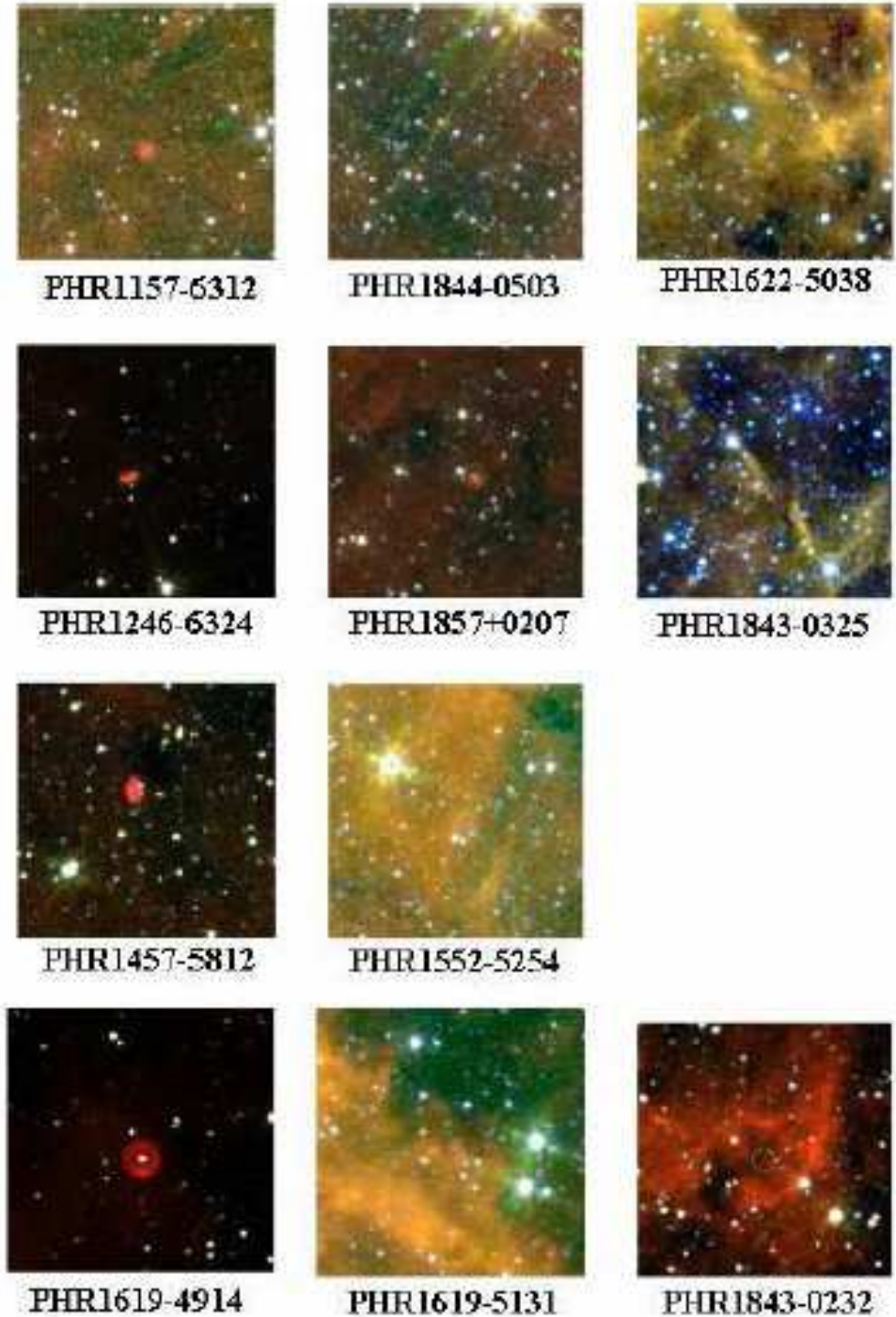} %pap3tif.eps} 
\caption{False color images of PNe from GLIMPSE mosaics (not residual images).  
PHR designations appear below each PN image.
In the bottom right corner is the field of PHR1843$-$0232, illustrating graphically the extreme
difficulty of extracting any object in the green circle against the very bright and variable 8.0-$\mu$m
sky background (red) emission of PAHs. All images are of 4\,arcmin square fields and any small green 
circle that guides the eye to the location of a PN's optical centroid has a 12$^{\prime\prime}$ radius.  \label{3color}}
\end{figure}

We intended the 3-color image to be a simple pragmatic tool that treats every PN's
trio of images identically.  We used SAOImage {\sc ds9} in its RGB mode.  Each of the three color
images for a given PN was displayed with the same stretch: linear and zscale.  
These images were examined for evidence of an extended counterpart to each PN, that had a
distinctive color compared with  its surroundings. 
 Clearly, MIR-bright PNe will stand out against a dark sky background, whatever their false color
(typically red).  Fig.~\ref{3color} displays 10
PNe that are recognizable by this technique.  Six are very obvious examples demonstrating the power
of this identification technique, while 4 others illustrate its capability for robustness in the presence 
of complicated MIR background emission.  Indeed, we are investigating the prospects for discovering new PNe, too
obscured to be visible in the SHS, by using GLIMPSE 3-color imaging alone.

PHR1844$-$0503 (top image of second column of MIR composite images in Fig.~\ref{3color}), whose position is
indicated by the small green circle based on the optical position, 
epitomizes the obvious matches, despite confusion with a diffraction spike
from a bright source at the northern edge of the image.  The orange periphery and yellow core 
distinguish the PN from artifacts and from the many stars seen in the field.

PHR1157$-$6312 (top image of first column in Fig.~\ref{3color})
lies in a field that is relatively sparse in stars but suffers from bright MIR ``cirrus'' emission 
that permeates almost the entire 4$\times$4\,arcmin field that we show.  The PN is the scarlet diffuse patch 
just below the field center, strongly contrasting with the widespread orange diffuse emission 
and predominantly green noise.  Twenty one nebulae (36\% of the 58) have definite, resolved, counterparts that are 
distinguishable from their environs by false color imaging. The dominant colors associated with these 
21 MIR counterparts separate into three groups: %8 are red (38\%); 10 are violet (48\%); and 3 are orange (14\%).  
8 are red; 10 are violet; and 3 are orange.  

Red most likely represents dominant 7.7 and 8.7-$\mu$m PAH band emission.  Orange could imply either of   
two explanations: PAH emission with strong 6.2-$\mu$m emission 
(comparable intensities in the 5.8 and
8.0-$\mu$m bands would lead to orange or yellow in false color) or H$_2$ emission lines which are strongest in   
IRAC's 8.0-$\mu$m band but also contribute to the 5.8-$\mu$m band chiefly through the 0$-$0\,S(7) line
(e.g. Hora et al. 2006). Violet objects could also represent two types of PN. In high-excitation PNe,  
ionic lines of heavy elements such as [Mg{\sc iv}] (4.485~$\mu$m) and [Ar{\sc vi}] (4.530~$\mu$m)
are very strong and fall in IRAC's 4.5-$\mu$m band.  However, such an object would also emit strongly
in the 8.99-$\mu$m [Ar{\sc iii}], 7.90-$\mu$m [Ar{\sc v}], and 7.65-$\mu$m [Ne{\sc vi}] lines in the
8.0-$\mu$m band. A combination of high-excitation lines would increase emission in both the
4.5-$\mu$m and 8.0-$\mu$m bands, leading to violet in the false-color image.
As a second scenario we suggest another possibility, namely low-excitation PNe in
which the H recombination lines are the dominant spectral features.  The combination of
Br$\alpha$ and Pf$\beta$ (in the 4.5-$\mu$m band) would overwhelm Pf$\alpha$ and lines such as
H{\sc i} 6$-$8 (in the 5.8-$\mu$m band).  Flux density ratios 4.5-$\mu$m/8.0-$\mu$m in typical PNe are 
very small except when H lines are dominant.  Ratios of 5.8-$\mu$m/8.0-$\mu$m emission
without substantial extinction, for PNe dominated by (i) PAH bands, or (ii) H$_2$ lines, or (iii) H recombination 
would be about 0.5 (Cohen \& Barlow 2005, Tables~2,3), 0.3 (Hora et al. 2006, Table~2), 
and 0.1 (using the H recombination lines in IC\,418: Pottasch et al.\,2004), respectively.
Bernard-Salas \& Pottasch 2001), respectively.  

Although the statistics are very limited, we note from the available MASH optical
spectroscopy that the average ratio of [N{\sc ii}]/H$\alpha$
for the seven PNe with red 3-color images (excluding PM\,5 which has He{\sc ii} rather than H$\alpha$) 
is 2.4$\pm$0.5, the three orange PNe give 6.3$\pm$1.7, while that for ten nebulae with violet
images is 1.0$\pm$0.5.  It appears that the false color combination of IRAC bands 2, 3 and 4
can differentiate between PNe of high and low [N{\sc ii}]/H$\alpha$.  Thus we propose the following
model to explain the MIR false colors and optical excitations of PNe.  ``Red'' PNe are PAH-dominated
and of modest excitation, ``orange'' objects are dominated by H$_2$ or high-excitation fine-structure
lines and ``violet'' nebulae are dominated by H recombination lines and are of low or very low excitation.

For the 22 PNe with [N{\sc ii}]/H$\alpha$ $\geq$3, 9 objects (41\%) have a bipolar morphology, suggesting that this line
ratio may be a valuable proxy for selecting Type~I PNe (see also Parker et al., in preparation). 
No such patterns emerge in the smaller sample of PNe with measured ratios of [O{\sc iii}]/H$\beta$ intensities.

\subsection{Overlays of MIR on H$\alpha$ images}
To locate the MIR counterparts the H$\alpha$ images were regridded to
Galactic coordinates with the same projections as the GLIMPSE maps and 
each of the four IRAC images was overlaid as contours on the H$\alpha$ image. 
``Quartets'' of these overlays were 
inspected.  Most of these IRAC counterparts are clearly resolved and many have
similar morphologies to that shown by the PNe in H$\alpha$, at least at the shorter IRAC wavelengths.
The lowest contours displayed in all quartets were set as the mean off-source
level of emission plus 3 standard deviations.
Filaments of PAH emission (in
the 5.8 and 8.0-$\mu$m bands) that might represent limb-brightened PDRs
will appear displaced from the periphery of a PN. In addition, one can 
readily recognize potential MIR counterparts to the PN central stars. 
For 9 of the 58 PNe (16\%) we could identify a candidate central star in one or more IRAC bands.  
Generally these are blue; i.e. they become monotonically fainter
with increasing IRAC wavelength.  

To illustrate how the quartets were used to determine the MIR counterparts
to PNe, we present four such quartets, chosen to highlight different
features in the nebulae, their probable MIR emission processes, and their central stars.  

\begin{figure}
\includegraphics[scale=.70]{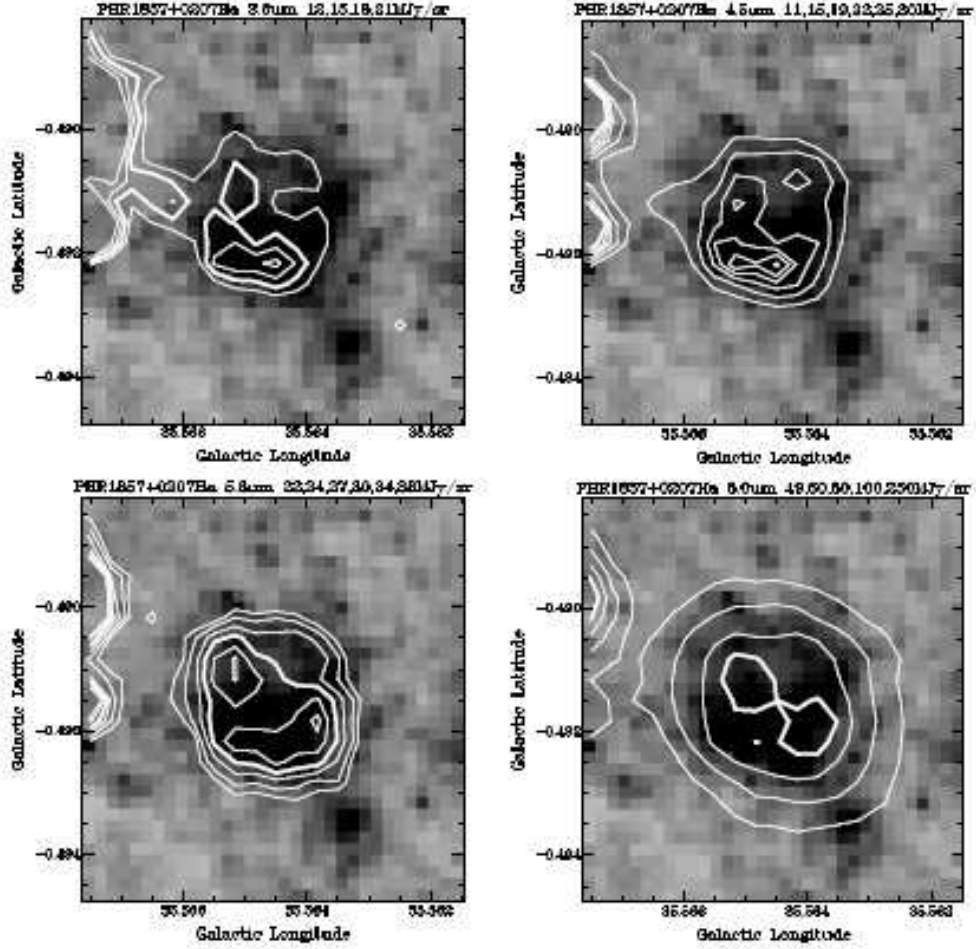} %f03557m049nall.epsi} %preprint
%\plotone{f11.eps}  %2column
\caption{Quartet of IRAC contours (in white) over the grayscale H$\alpha$ image 
of PHR1857+0207. The MIR counterpart of this PN increases in size with increasing wavelength,
perhaps indicative of emission by fine-structure lines in the 3.6 and 4.5-$\mu$m bands, and 
PAHs at 5.8 and 8.0\,$\mu$m. Contour levels at 3.6, 4.5, 6.8, 8.0\,$\mu$m are: 12, 15, 18, 21\,MJy\,sr$^{-1}$;
11, 15, 19, 22, 25, 30; 22, 24, 27, 30, 34, 38; and 49, 60, 80, 100, 250,
respectively.  \label{qtet1}}
\end{figure}

The first quartet, presented in 
Fig.\ref{qtet1}, shows the small nebula PHR1857+0207 that appears in an otherwise empty field.   
Note the progressive increase in size of this likely PN with increasing wavelength, 
and the change in its morphology from the asymmetric enhanced brightness of the 
southern limb in the three shortest bands to the large circular appearance
at 8.0\,$\mu$m.  We identify the 8.0-$\mu$m structure and the increase in size
at 5.8\,$\mu$m as due to a substantial PDR that envelops the entire ionized 
zone.  Ionic fine structure lines probably account for the changing
size from 3.6 to 5.8\,$\mu$m.

\begin{figure}
\includegraphics[scale=.70]{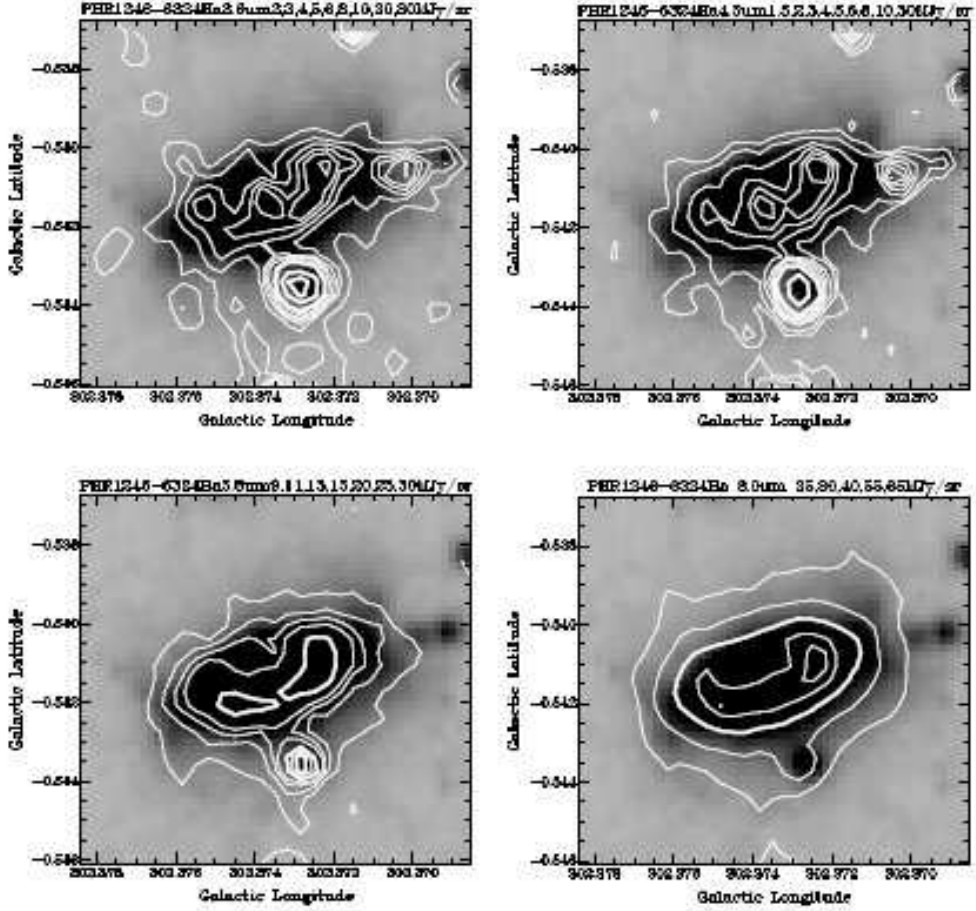} %f30237m054nall.epsi} %preprint
%\plotone{f12.eps} %2column
\caption{Quartet of IRAC bands overlaid on H$\alpha$ for PHR1246$-$6324.  The MIR images suggest
dominantly thermal emission by warm dust in a tilted disk, although the greater extent at
8.0\,$\mu$m might also signify a PDR.
Contour levels at 3.6, 4.5, 6.8, 8.0\,$\mu$m are: 2, 3, 4, 5, 6, 8, 10, 20, 30\,MJy\,sr$^{-1}$;
1.5, 2, 3, 4, 5, 6, 8, 10, 30; 9, 11, 13, 15, 20, 25, 30; and 25, 30, 40, 55, 65,
respectively. \label{qtet2}}
\end{figure}

The second quartet, Fig.~\ref{qtet2},
 offers an optically and MIR-bright true PN, PHR1246$-$6324.
This field has several blue stars at 3.6 and 4.5\,$\mu$m, in particular 
one bright star just below the PN's center, which seems to be unrelated 
to the nebula.  The number of stars detected falls with
increasing wavelength (as expected for random stars in their Rayleigh-Jeans
domain) and only the brightest one is still detected at 8.0\,$\mu$m. The
PN is bipolar and the bright H$\alpha$ axis is filled across the pinched waist.  This
elongation is also seen in the MIR.  Unlike in Fig.~\ref{qtet1}, the 3.6 and 4.5-$\mu$m
bands are almost identical in appearance and dimensions,  which closely match the
distribution of ionized gas traced by H$\alpha$.  This is due either to recombination 
lines in these two short bands (e.g. Pf$\gamma$ and Br$\alpha$, 
respectively) or to thermal emission from dust grains, close to the central
star, which are heated by direct starlight.  In both 
the longer bands the PN is markedly larger in latitude extent, reflecting
the contribution by PAHs in the PDR in the outer portions of the nebula. 
The inner contours suggest thermal emission 
by cooler grains in a somewhat tilted dust disk.  This dust emission must be
optically thin to produce the two peaks, presumably from limb-brightening
along the line-of-sight to the circumstellar dust disk.  The U-shaped
curvature of the second highest 8.0-$\mu$m contours strongly suggests
that IRAC has resolved a tilted dusty disk.  

\begin{figure}
\includegraphics[scale=.70]{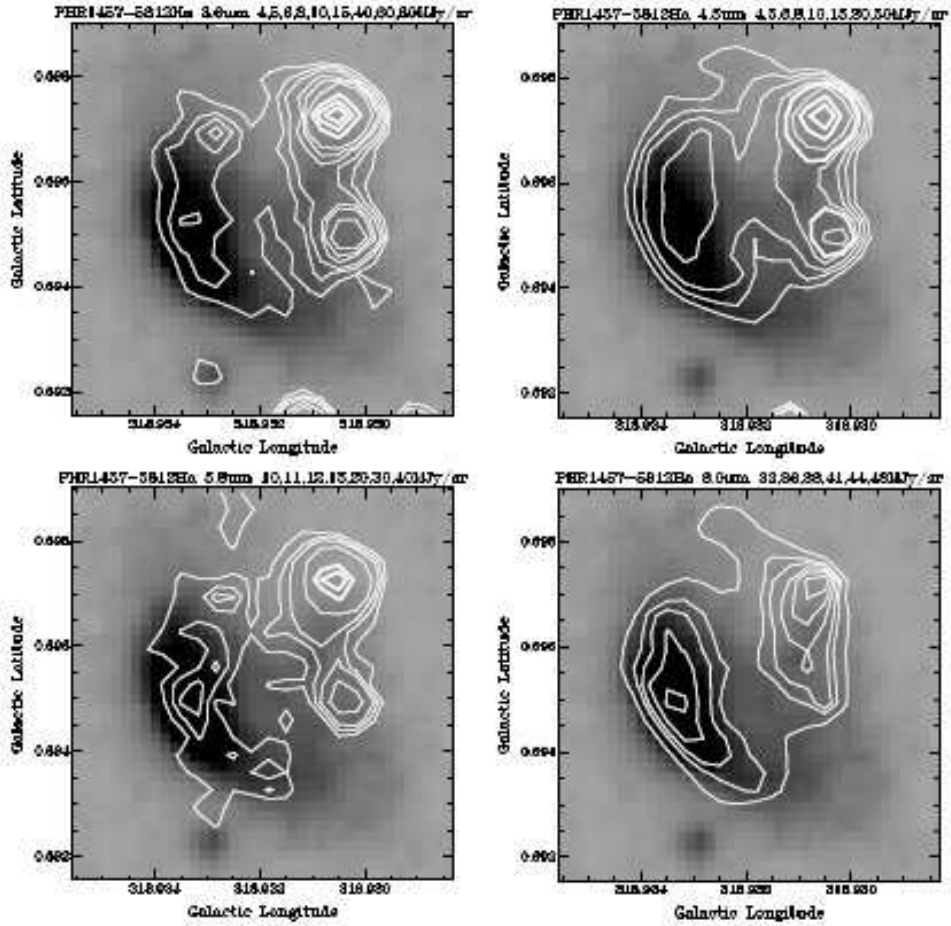} %f31893p069nall.epsi} %preprint
%\plotone{f13.eps} %2column 
\caption{Quartet of IRAC bands overlaid on H$\alpha$ for PHR1457$-$5812.  We suggest the MIR
structure is due to H$_2$ line emission.
Contour levels at 3.6, 4.5, 5.8, 8.0\,$\mu$m are: 4, 5, 6, 8, 10, 15, 40, 60, 80\,MJy\,sr$^{-1}$;
4, 5, 6, 8, 10, 15, 30, 50; 10, 11, 12, 15, 20, 30, 40; and 32, 36, 38, 41,
44, 48, respectively. \label{qtet3}}
\end{figure}

Fig.~\ref{qtet3} presents PHR1457$-$5812, in the third quartet;
a true PN with a very different MIR 
structure, also shown in Fig.~\ref{8umha} and in Fig.~\ref{3color} (left column, third image down).  
One must ignore the two unrelated point sources projected against the 
western edge of the PN. It is a compact but asymmetric PN with a strongly enhanced eastern
edge. 
The curving eastern portion
with the bright H$\alpha$ emission coincides with the peaks of MIR 
counterparts in the three longer IRAC bands.  These MIR peaks shift slightly 
with wavelength but the spatial extent of
the resolved structure is remarkably similar at 3.6 and 5.8\,$\mu$m where
it lies inside the H$\alpha$ image, and again at 4.5 and 8.0\,$\mu$m, where
it is slightly more extensive with a boundary very well-matched to that of 
the ionized gas.  Despite the redness (Fig.~\ref{3color})
this pattern does not suggest the presence of PAHs.  It seems more
indicative of strong emission by the pure ($\nu$=0-0) rotational lines of 
H$_2$ within the IRAC bands.  These lines dominate the MIR emission of the
Helix Nebula (Cox et al. (1998); Hora et al. (2006)) and are strong in 
high-excitation PNe (Bernard-Salas \& Tielens 2005).  This would 
also account for the nebular detection by 2MASS, which is brightest in the 
K$_s$ band that includes H$_2$ from the strong S(1) 1-0 line. 
Optical spectra available from MASH for this object show [N{\sc ii}]/H$\alpha$ $\sim$4 
but only weak [O{\sc iii}] in the blue due to high extinction.
There is also weak [Ar{\sc iii}] at 7136\AA\, but no evidence of higher excitation He{\sc ii}.

\begin{figure}
\includegraphics[scale=.70]{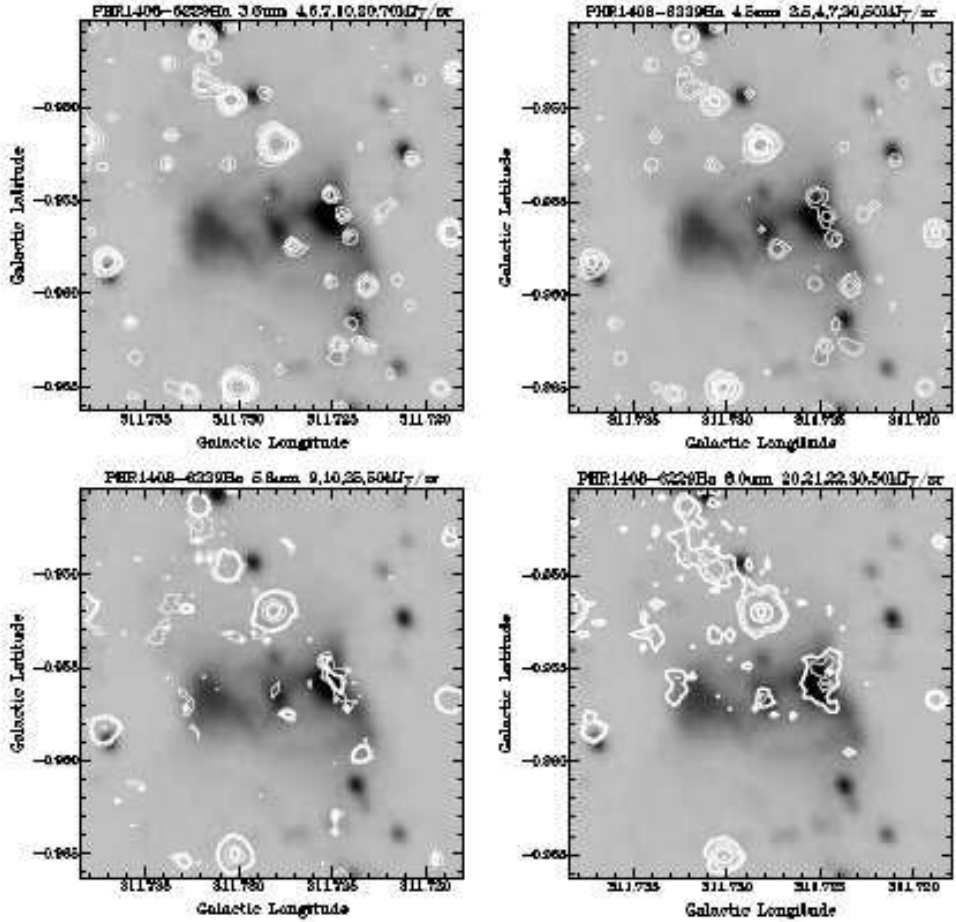} %f31173m095nall.epsi} %preprint
%\plotone{f14.eps} %f31173m095nall.epsi} %2column
\caption{Quartet of IRAC bands overlaid on H$\alpha$ for PHR1408$-$6229.  This large high-excitation 
bipolar PN shows a diffuse waist terminated to east and west by bright MIR PDRs.
Contour levels at 3.6, 4.5, 6.8, 8.0\,$\mu$m are: 4, 6, 7, 10, 20, 70\,MJy\,sr$^{-1}$;
2.5, 4, 7, 20, 50; 9, 10, 25, 50; and 20, 21, 22, 30, 50, respectively.  \label{qtet4}}
\end{figure}

The fourth quartet appears in Fig.~\ref{qtet4}, which 
represents the inner parts of the bipolar true
PN PHR1408$-$6229, whose brightest H$\alpha$ component is the east-west
edge-on disk that is shown in each figure.  The two flanking regions 
represent optically-thin, limb-brightened, 
ionized gas viewed at the extreme edges of a circumstellar torus of gas. 
The bipolar lobes are much fainter and extend far to the north and south of this disk.  
Three stars appear against
the western side of the torus at 3.6 and 4.5\,$\mu$m but disappear at 5.8\,$\mu$m, to be replaced by a small
diffuse patch.  This becomes more prominent at 8.0\,$\mu$m.  We interpret this, and its fainter
eastern counterpart, as PDRs lying just outside the bright ionized edges of the central
disk. The peak of the eastern PDR is clearly displaced from the H$\alpha$ emission.  The PN has
a size in H$\alpha$ of 82$^{\prime\prime}$$\times$46$^{\prime\prime}$.  
Large PNe lack PAHs, perhaps
because these lower density nebulae are optically thin to ionizing radiation and have no PDRs.
The only large PNe in which PAHs have been detected are bipolar, high-excitation PNe in which
PAHs are found in a high-density central circumstellar disk.  We note that PHR1408$-$6229 has
[N{\sc ii}]/H$\alpha$ of 8, and shows [O{\sc iii}] $>$ H$\beta$ so it may well be another 
high-excitation bipolar PN.  It is fortunate that a blue spectrum of this
object exists to support this conclusion.  Often the MASH PNe have lower
quality blue spectra because the heavy extinction so close to the Plane
requires extremely long integrations to detect a useful result.

\subsection{The fraction of PNe with the most convincing MIR counterparts}
Not every PN that can be recognized by false-color
imaging can be meaningfully extracted from its surroundings; for example, 
PHR1843$-$0325 (Fig.~\ref{3color}; bottom right corner).
We focus now on the most convincing MIR counterparts of MASH nebulae.  These
are PNe for which robust quantitative estimates can be made of 8-$\mu$m emission
above their surroundings.  Among the 58 PNe we found 14 (24\%) such MSX 
counterparts and 19 (33\%) using GLIMPSE images due to the higher sensitivity 
and resolution of IRAC.  
These seem like small numbers of MIR counterparts detected
compared to the success rates enjoyed by the MIR surveys made in the 1980s of
the then known PNe, later compiled into the highly heterogeneous catalogs of
Acker et al. (1992,1996) and Kohoutek (2000).  What differentiates the MASH PNe 
from the traditional, but highly heterogeneous compilations of PNe first assembled at Strasbourg?

Firstly, the previously known PNe catalogued by Acker and colleagues and
independently by the various compilations of Perek and Kohoutek (e.g. Kohoutek 2000),
were largely detected originally in the optical from broadband, narrow-band, or
objective-prism photography. Unsurprisingly, these samples represent the bright end of the
PN luminosity function and those nearest the sun which are less extinguished.
It is only recently, with the advent of new surveys in the infrared and
optical using, for example, combinations of selected narrow-band filters and CCD imaging, that
more extinguished, lower surface brightness and more evolved PNe have been detected in significant
numbers. A brief summary of these small-scale projects appears in the main MASH paper of
Parker et al. (2006) but it is the MASH survey itself that has changed the situation most
dramatically (see later).

Early MIR work drew upon the previously published PN compilations and found all the IR-bright 
nebulae amongst the known optical PNe.  For example, at 10\,$\mu$m Cohen \& Barlow (1974,1980)  
detected 52\% of their combined total of 145 of the optically best-known PNe with apparent 
diameters $\leq$30$^{\prime\prime}$.  Consequently, a large fraction would be expected to have MIR
counterparts. Cohen \& Barlow (1974) also noted that those PNe whose central
star showed an optical emission-line spectrum were more likely to have MIR detections. 
Other programs have made use of the specific $IRAS$ colors of PN candidates through the 
far-infrared (60 and 100\,$\mu$m) excess associated with the cool dust in the nebulae.
Ratag \& Pottasch (1991) and van de Steene \& Pottasch (1995) identified 63 new PNe and 67
possible PNe. There is always a question mark over such candidates until confirmatory 
spectroscopy and high resolution optical, NIR, or radio imaging are obtained.

Young compact PNe are dense, surrounded 
by dust grains that are heated dominantly by direct starlight.  Most of 
these were already known from early optical searches.  Mature PNe that
have expanded significantly derive most of their
MIR thermal emission by absorbing resonantly trapped Ly-$\alpha$ photons 
(Cohen \& Barlow 1974). This mechanism acts as a thermostat so that all 
grains in the ionized zone of a PN see approximately the same
intensity of UV radiation and attain the same temperature, typically around
120\,K.  At this temperature, the bulk of the thermal dust emission 
is radiated at wavelengths longer than 25\,$\mu$m and does not
lead to bright emission near 8\,$\mu$m from PNe.  The dominant
emission processes that can be observed in PNe by IRAC are fluorescent
PAH bands from the PDR, fine structure lines from the ionized zone, 
stratified outward in the nebula according to decreasing excitation level,
and H$_2$ lines that arise chiefly in the warm regions of the PDR.
Forbidden atomic lines also emit in the PDR but these radiate principally
between 35 and 158\,$\mu$m, outside the range of the IRAC.

MASH has greatly increased the number of the most highly evolved optical
PNe known.  Indeed, it has discovered even more extreme examples of the
phenomenon of PNe that are dissolving into the interstellar medium, e.g. Pierce et al. (2004).
Such PNe are inherently of low surface brightness, except at possible shock fronts, and so become undetectable at
increasing distance from the sun and as the interstellar extinction levels become significant. MASH PNe span a broader
evolutionary range than the previously published PNe and are generally more evolved, obscured,
of lower surface brightness and greater angular extent than those of most other PN catalogs. These combined
properties naturally lead to lower levels of MIR detectability with current survey sensitivities including   
GLIMPSE, though the success rate here is higher than for the same region covered by MSX. These characteristics and
the implications for MIR detectability are explored in more detail below.

\subsection{The MIR attributes of MASH PNe}
As a PN ages many factors affect its MIR-detectability.  When the density drops below 
N$_e$$\sim$1000\,cm$^{-3}$, Ly-$\alpha$ photons are no longer trapped in the ionized zone
and dust grains can absorb only dilute starlight that provides little thermal heating, 
particularly as the central star descends the white dwarf cooling track.
Lower density nebulae are optically thin to ionizing radiation and have no PDRs unless they
are also bipolar and surrounded by dense circumstellar disks (Cohen \& Barlow 2005).
Extensive neutral and molecular material resides outside many PNe from previous mass lost
while the star was on the asymptotic giant branch (AGB).  Indeed, the bulk of the stellar
ejecta in PNe must occupy extensive, but optically-faint, AGB halos that are created by slow
AGB winds and are either neutral or only weakly ionized.  
Advancing age also increases the fraction of total PN mass in the form of ionic gas, while
decreasing the atomic and molecular gas masses (Bernard-Salas \& Tielens 2005), reducing
PAH and H$_2$ emission and lessening nebular detectability by IRAC.   While these authors
emphasize the importance of the PDRs around PNe they state explicitly that these regions
are thin compared with the ionized zones.  Therefore, they do not treat extended halos in
their analysis so their ``total mass" excludes AGB halo material.  However, their overall results 
are robust as regards the
PDRs and the ionized regions of these nebulae to which the MASH, IRAC, and radio continuum are sensitive.
As a nebula ages and expands, the radiation field at large radii is more dilute even as the nebula itself
becomes optically thin.  Those PAHs that still survive are bathed in a diminishing far-UV radiation field,  which can be
quantified by
plotting incident far-UV flux against PN diameter (Bernard-Salas \& Tielens 2005: their Fig.~3).  
Recombination occurs if denser clumps remain in the inner nebula and previously neutral gas continues
to expand slowly outwards.  

Not every PN is carbon-rich, containing PAHs.  Oxygen-rich PNe also have
dust grains, identified by their silicate spectral features, which peak
near 10\,$\mu$m.  These contribute relatively little to the IRAC
8.0-$\mu$m band, whose relative spectral response curve drops abruptly
beyond 9\,$\mu$m.  Furthermore, for a PN with a low gas-phase ratio of C/O,
the fraction of total nebular IR luminosity emitted in PAHs is
correspondingly decreased (Cohen \& Barlow 2005: their Fig.~3).  Therefore,
fewer 8-$\mu$m counterparts should be found among PNe that are highly evolved, 
or physically large, or of low C/O abundance ratio,  
and among those for which the central stars exhibit no stellar winds
(seen via Wolf-Rayet or Of emission lines).  Of our 58 PNe, only PM\,5 has a 
known emission-line central star. 
Consequently, a much smaller proportion of MASH PNe are expected to have
MIR counterparts, as has been found.

Preliminary central star identifications have been made for about 
15\%-20\% of the MASH catalog nebulae, mostly based on a method in which
all the available photometric images (SuperCOSMOS Sky Survey, H$\alpha$, blue and red 
images, etc.) have been examined.  Central (blue) star \emph{candidates} were identified
by combining the UKST $B_j$ (IIIaJ), UKST Short Red, and H$\alpha$, 
images as the blue, green and red elements of a false-color
composite and blinking this composite with differences or ratios between 
$B_j$ and $R$ (UKST Red, IIIaF) images.  On close inspection,
many of the fainter stars in these tri-color composites are ``blueish", a
consequence of the fainter limiting magnitude of the blue IIIaJ emulsion
in general.  In a crowded field there is a good chance of one of these
faint ``blue" stars appearing close to the geometric center of an
extended PN and being identified as the true CS.  $UBVI$ photometry is,
therefore, needed in all doubtful cases.  However, because there are so few
candidate blue stars in this paper (most PNe in GLIMPSE are at very low-latitudes 
and suffer high extinction), it is not really a factor.  PHR1447-5838 and PHR1244$-$6231 
are currently the only really clear cases of a blue CS candidate.

\subsection{ Notes on individual objects}
Individual MASH PNe are detailed below if they illustrate aspects of the process of establishing
multiwavelength identifications.  

\noindent
PHR1813$-$1543: the radio counterpart appears to be a double source, perhaps
a background radio galaxy.  There is no MIR counterpart within the PN nor around its 
periphery.

\noindent
PHR1826$-$0953: The radio source is located exactly at the H$\alpha$ 
  centroid of the PN. The MSX counterpart of a bright star fills much of
  the area of the PN. In the IRAC mosaic image the stellar diffraction
  vanes make accurate analysis impossible. However, the GLIMPSE residual images indicate 
that the star is not associated with the PN because it lies far from the optical centroid
of this oval bipolar nebula.
Bright diffuse emission lies across the PN so that it is difficult 
  to provide a useful upper limit with MSX ($<$160\,mJy at 8.3\,$\mu$m is 
  given in Table~2). There is clearly an excess of MIR emission within the
  PN and this can be measured using the residual images to give an IRAC 
  8.0-$\mu$m detection of 170\,mJy.

\noindent
PHR1843$-$0232: The NVSS radio source in this vicinity is not associated with the PN.
There are clear indications of a surrounding PDR that wraps around 270$^\circ$ of
the PN's outer rim at 8.0\,$\mu$m.  However, the region is suffused by bright streamers and
extended 8.0-$\mu$m emission  so that an estimate of the MIR emission from
the PDR is impossible. This situation occurs frequently and emphasizes the 
impossibility of undertaking any uniformly flux-limited survey within
the Galactic plane. The problem is illustrated for this PN in the final panel of the
montage of 3-band false-color images in Fig.~\ref{3color} (bottom right
corner). The 8-$\mu$m sky brightness in the immediate vicinity of the PN
is about 70\,MJy\,sr$^{-1}$. 

\noindent
PHR1457$-$5812: The MSX Point Source Catalog ver.2.3 (PSC2.3: Egan et al. 2003) offers only
an upper limit at 8.3\,$\mu$m of $<$110\,mJy, consistent with our detection 
of a spatial integral of 78\,mJy above local background in a 
region encompassing the PN's H$\alpha$ extent.  

\noindent
PHR1619$-$4914: This object is ``PM\,5", the only PN known in the Galaxy
to have a central Wolf-Rayet star of type [WN] (Morgan, Parker, \& Cohen 
2003). Although the MSX PSC2.3 lists a value of 600\,mJy at 8.3\,$\mu$m
this is not of the entire PN, which is
substantially larger than the MSX point spread function (PSF), nor does it reflect an
accurate estimate of the CS which lies in bright PN nebulosity. An estimate
using the IRAC images suggests that the CS contributes about 25\% of the PN's total 
integrated flux at 8.0\,$\mu$m.  A more detailed examination of this PN with {\it Spitzer}
will be presented by Cohen, Shupe, \& Parker (in preparation).

\noindent
PHR1223$-$6236: This object shows weak evidence for an association between the PN and diffuse
patches of 5.8 and 8.0-$\mu$m emission on the NE and SW rims.  A bright star is
projected against the NE rim making it difficult to distinguish between artifacts
of this point source and MIR extended emission.  However, the residual
image definitively shows diffuse 8.0-$\mu$m emission distinct from the stellar PSF.

\begin{figure}
\includegraphics[scale=.70]{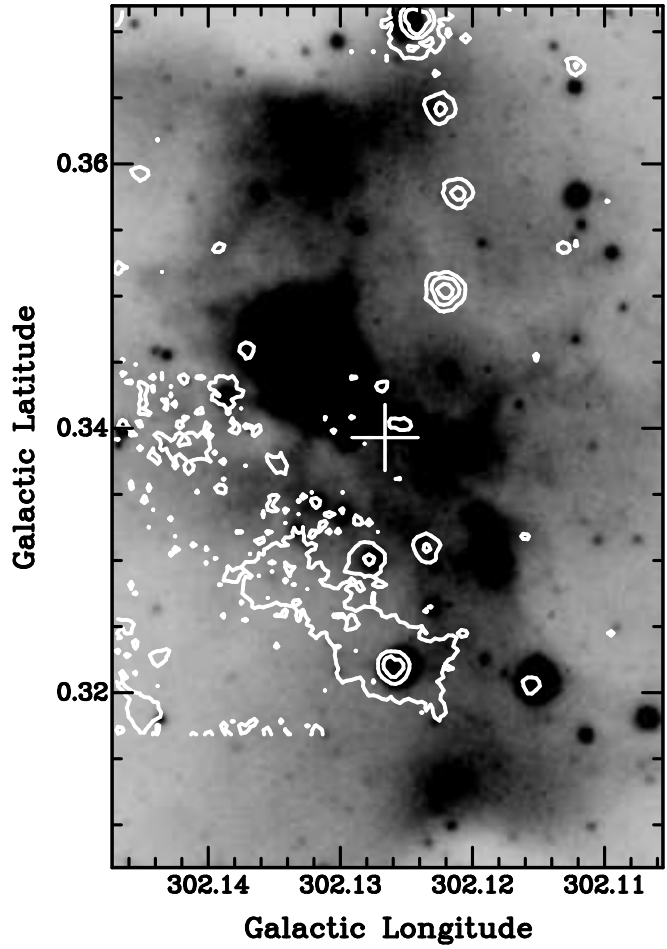}  %{rcwha_8.eps} %preprint
%\plotone{f15.eps} %2column
\caption{White contours of 8.0-$\mu$m emission (20, 40, 100\,MJy\,sr$^{-1}$) overlaid
on the gresycale H$\alpha$ image of PHR1244$-$6231.  The horizontal line of 8.0-$\mu$m
patches near 0.317 longitude is an artifact due to a very bright star outside the field
presented, to the south-east. \label{rcwha}}.
\end{figure}

\noindent
RCW\,69 (PHR1244$-$6231) is a relatively nearby, evolved Type~I bipolar PN
(Frew, Parker, \& Russeil 2006), viewed through the Coalsack.  RCW\,69 is one of the closest 
PNe in the GLIMPSE sample (1.3$\pm$0.2\,kpc) and it is also intrinsically large 
(1.6$\times$1.5 pc$^{2}$).  It has an elongated central bar in optical images which likely 
represents a thick edge-on torus similar to the ring seen in the Helix nebula (O'Dell,
McCullough, \& Meixner 2004), but it is much fainter in H$\alpha$ surface brightness than the
Helix and is more evolved.  Frew et al. (2006) identify the $B$=18.4 CS and 
demonstrate that it is on the white dwarf cooling track.  This star is undetected in GLIMPSE images.
Fig.~\ref{rcwha} overlays 8.0-$\mu$m contours on an H$\alpha$ greyscale image of the PN.
The location of the CS is marked by a cross.  There is evidence for a PDR in the form of 
a bar of 8.0-$\mu$m emission displaced to the east of the obvious nebular bar, making RCW\,69 
another example of a large bipolar PN associated with PAH emission.  Frew et al. (2006) 
discuss the possible detection of CO\,(1-0) and CO\,(2-1) emission which indicates the presence 
of molecular material in the PN although its location is unknown.

\noindent
PHR1250$-$6346: The association (in Table~\ref{pnelist}) between MIR emission
and this moderately large PN  seems likely because
there are indications of a peripheral PDR around the western rim in the
form of diffuse emission at 5.8 and especially at 8.0\,$\mu$m.

\section{The diffuse calibration of IRAC}
Cohen et al. (2006) have recently examined the diffuse 8.0-$\mu$m calibration of IRAC using a sample
of 43 H{\sc ii} regions observed by MSX and by the {\it Spitzer} Telescope. For an
angular spatial scale of up to
$24^{\prime}$ they found a median ratio of IRAC 8.0-$\mu$m to MSX 8.3-$\mu$m spatially
integrated fluxes of 1.55$\pm$0.15.  A factor of 1.14 is caused by the different contributions
made by PAH emission in the two very different space-based bandpasses.  The remaining 
component corresponds to an overestimate by 36\% in the instrumental calibration of IRAC at
8.0\,$\mu$m, as was independently noted by Reach et al. (2005).
The {\it Spitzer} Science Center (SSC) has
recommended that extended source 8.0-$\mu$m fluxes be scaled down by a factor of 0.74
(see also comparable work based on the light distributions of elliptical galaxies 
\footnote{http://ssc.spitzer.caltech.edu/irac/calib/extcal/}). 

The MASH PNe have been used to check the absolute diffuse calibration of IRAC at 
8.0\,$\mu$m on smaller spatial scales than in our study of H{\sc ii} regions, spanning 
the range from PNe only slightly larger then the IRAC PSFs (but smaller than the MSX PSFs) to a 
scale of a few arcmin. Details of the
method adopted to calculate directly comparable integrated fluxes from the MSX images of the PNe 
and the GLIMPSE residual images are given by Cohen et al. (2006).  To represent the
PNe we utilized the H$\alpha$ images from the SHS (Parker et al.
2005).  The outer H$\alpha$ contours of each PN were overlaid on its quartet of IRAC
images and the flux density integrated over the identical area for all bands, while
including any peripheral PDR apparent at 8.0\,$\mu$m.  The same
approach was used with MSX 8.3-$\mu$m images.  Multiple estimates were made of the sky
background for each PN.  For the measurements of sky background
for each PN, specific areas were selected from the IRAC 8.0-$\mu$m residual
images.  The same areas were used for the three other bands. It must be
emphasized that it is critical to understand the contribution of the interstellar medium 
at 8.0\,$\mu$m in defining what is sky emission and what is a MIR counterpart of a PN.

Extraneous stars (not CS candidates) within the PN boundary were removed from the
MSX PN images using the MSX PSC2.3 and the Reject 
Catalog. For the study of H{\sc ii} regions  (Cohen et al. 2007) the areas
were as large as 1.5\,deg$^2$, and we used a validated statistical method for the
removal of contaminating point sources.  Many of these sources were
undetected by MSX with its lower sensitivity but were automatically removed
from the GLIMPSE residual images.  Therefore, in order to make a meaningful
comparison between IRAC and MSX spatially integrated fluxes, one needs to
subtract the contribution from the contaminants to the same depth for both
datasets. A more complete explanation is given in Cohen et al. (2007), section
 \S4.1.  The method utilizes the total surface brightness mode (Cohen 2001) of the ``SKY" model
for the point source sky (Wainscoat et al. 1992), operating in the 8.3-$\mu$m MSX band using its
embedded library of 2$-$35-$\mu$m archetypal spectra (Cohen 1993).  SKY calculates the diffuse sky surface 
brightness due to smearing of unresolved point sources.

The largest PN in our sample of 58 MASH objects (PHR1408$-$6106) encompasses an area of 18\,arcmin$^2$.
Within the H$\alpha$ boundary there
are two unresolved 8.3-$\mu$m sources from the MSX PSC2.3 catalog above the 5$\sigma$
level and none listed in the corresponding Reject Catalog.  The
background emission would permit detection of an 8.3-$\mu$m source at
magnitude 7.5 (60\,mJy).
SKY predicts that an additional 2\% of the sky background surface
brightness should be subtracted from the integrated PN flux to account for 
the sources (to a magnitude of 8.0) that were removed in producing the
GLIMPSE residual image in this field. Such a small adjustment is well within the uncertainties
of our MIR flux measurements.  The remainder of the MASH sample have smaller
areas and correspondingly smaller corrections to their observed 8.3-$\mu$m
integrated fluxes. Consequently, we have subtracted only those point
sources listed in the PSC2.3 and the Reject catalog to ensure equivalence
of the PN fluxes estimated from the MSX and GLIMPSE residual images. 

We also estimated upper limits for PNe undetected by MSX and/or IRAC as
three times the root-sum-squared 1$\sigma$ uncertainties in the fluxes measured
for both PN and sky background locations at each wavelength.   

Table~\ref{fluxes} summarizes the MIR integrated fluxes for 19 PNe. Objects
listed first have MIR detections from MSX and IRAC, together with a radio detection. Later entries
are PNe without radio detections but with both MSX and IRAC data. PNe with fewer than
two detections among MSX, IRAC, and the radio continuum are excluded.

\begin{figure}
\includegraphics[scale=0.60,angle=-90]{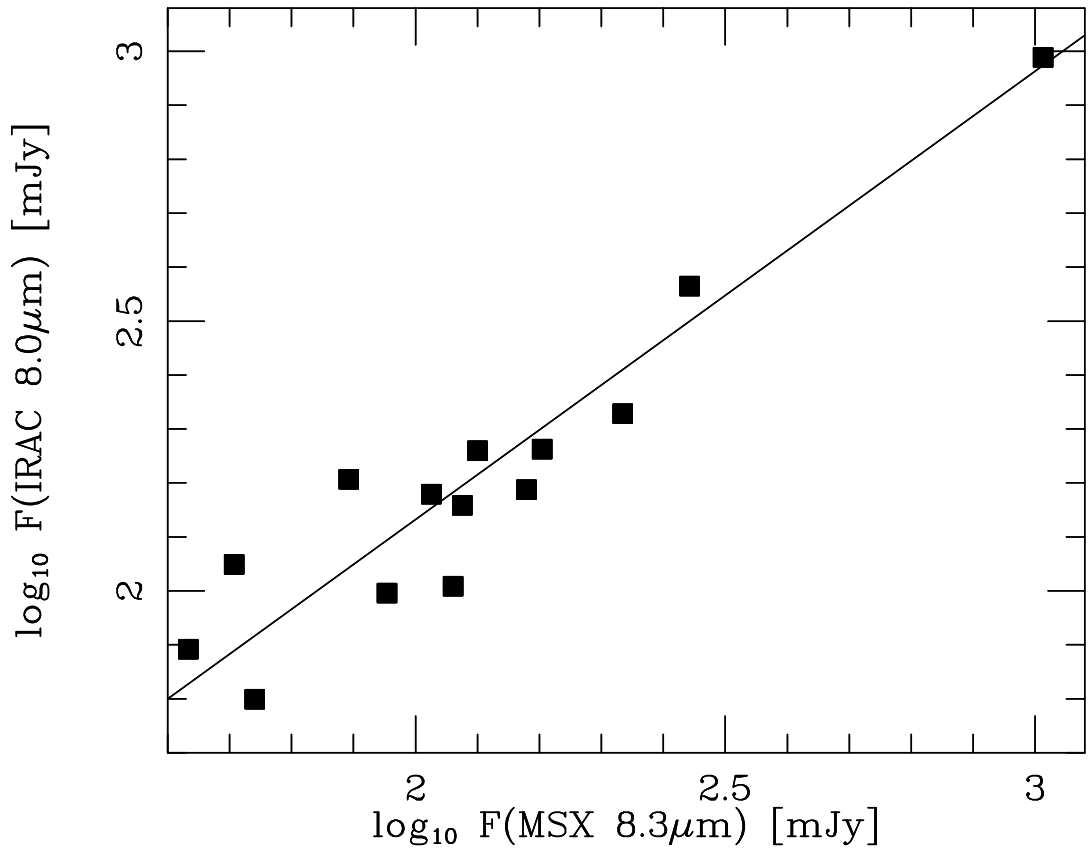} %pnsstmsx.eps}
\caption{Comparison of spatially integrated IRAC 8.0 and MSX 8.3-$\mu$m fluxes
for PNe.  The solid line is the regression based on errors in both variables.  \label{sstmsx}}
\end{figure}

A total of fourteen PNe have both
MSX and IRAC MIR detections.  This sample spans a dynamic range at 8\,$\mu$m from below
30\,mJy to about 1000\,mJy.  Fig.~\ref{sstmsx} compares these results and
plots a formal linear least squares regression line (with uncertainties
in both variables).  The slope of this logarithmic plot is 
0.9$\pm$0.1, consistent with a linear proportionality between IRAC and MSX fluxes.  The
offset is poorly defined as 0.4$\pm$0.2, corresponding to a ratio of IRAC/MSX of
2.9$\pm$1.5.  More accurate determinations come from the unweighted mean of the ratios of IRAC/MSX 
(1.3$\pm$0.2) and the median of the sample of fourteen PNe (1.2$\pm$0.2),
which are consistent with the factor of 1.55 found for this ratio from H{\sc ii} regions
(Cohen et al. 2007).  

\begin{figure}
\includegraphics[scale=0.60,angle=-90]{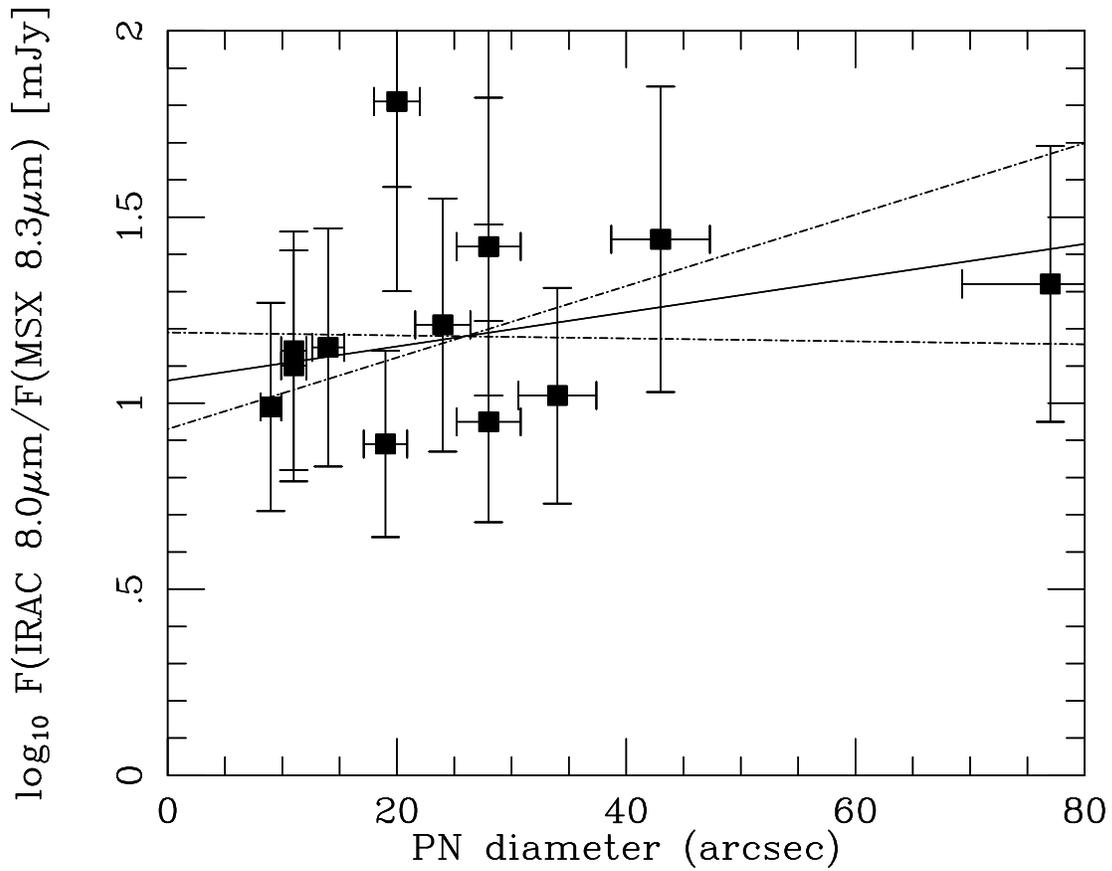} %pnratrade.eps}
\caption{Ratio of IRAC to MSX fluxes for PNe plotted against diameters
of the nebulae.  Solid line is the regression line flanked (dash-dotted) by the $\pm1\sigma$  
extreme fits.  Error bars are shown both for the ratios and for diameters. 
Note the slope of the regression line that suggests the correctness of the IRAC
calibration for a point source, and the steady increase in ratio to about 1.5
with increasing nebula size. \label{ratrad}}
\end{figure}

The fourteen PNe also span a wide range in spatial scale.  We used the geometric mean of
the MASH measurements of the major and minor axes of the nebulae in H$\alpha$ to calculate 
a representative ``diameter'', which varies from 9\arcsec\, to 77\arcsec.  Several
PNe are smaller than the MSX PSF and several are smaller in the MIR than in H$\alpha$ even
when resolved by IRAC.  A plot of the ratio of IRAC/MSX against
nebular radius will reveal if the calibration of IRAC diverges with increasing PN radius
from the expectation of a ratio of 1.0 at small scale (the point source calibration of
IRAC is good to $\sim$3\% absolute) to a value close to 1.55 for the largest scale at which we
can probe the diffuse calibration.   Fig.~\ref{ratrad} illustrates this comparison.

Three lines are overlaid on the plotted points: the formal regression line (solid) and
the $\pm1\sigma$ limiting relationships that couple the mean+1$\sigma$ slope with the
mean$-$1$\sigma$ offset, and vice versa (dashed lines).  The regression was
derived by assigning 10\% uncertainties to the diameters of the PNe, and using
the 1$\sigma$ errors in the ratios of MIR fluxes determined from the root-sum-squared errors
in both IRAC and MSX flux integrals.  The slope is 0.005$\pm$0.005, with an  offset of 1.1$\pm$0.1. 
The results are, therefore, marginal but would accommodate a gradual change between a point 
source calibration factor of 1.0 at small nebular size to a diffuse factor $\geq$\,1.4 by 77\arcsec.

\section{PN colors with IRAC}
Table~\ref{colors} summarizes the six color indices in Vega-based magnitudes derived from 41 of 
the 58 MASH PNe for which spatially integrated fluxes were estimated.  Not
every band was measureable above the sky for every nebula.  The median colors and the standard 
errors of the median (sem) are given after applying the SSC recommended aperture
correction factors in each IRAC band.  The color offsets that these cause are
given by Cohen et al. (2007: their Table~5, col.4).  For comparison we have synthesized  %(4))??
the same six colors for a set of 26 optically well-known PNe taken from Acker et al. (1992), 
based on low-resolution spectra obtained with the Short Wavelength Spectrometer (SWS)  of
the {\it Infrared Space Observatory} (ISO). There is excellent
agreement within the $\sim$1$\sigma$ joint uncertainties.

\begin{figure}
\includegraphics[scale=0.70]{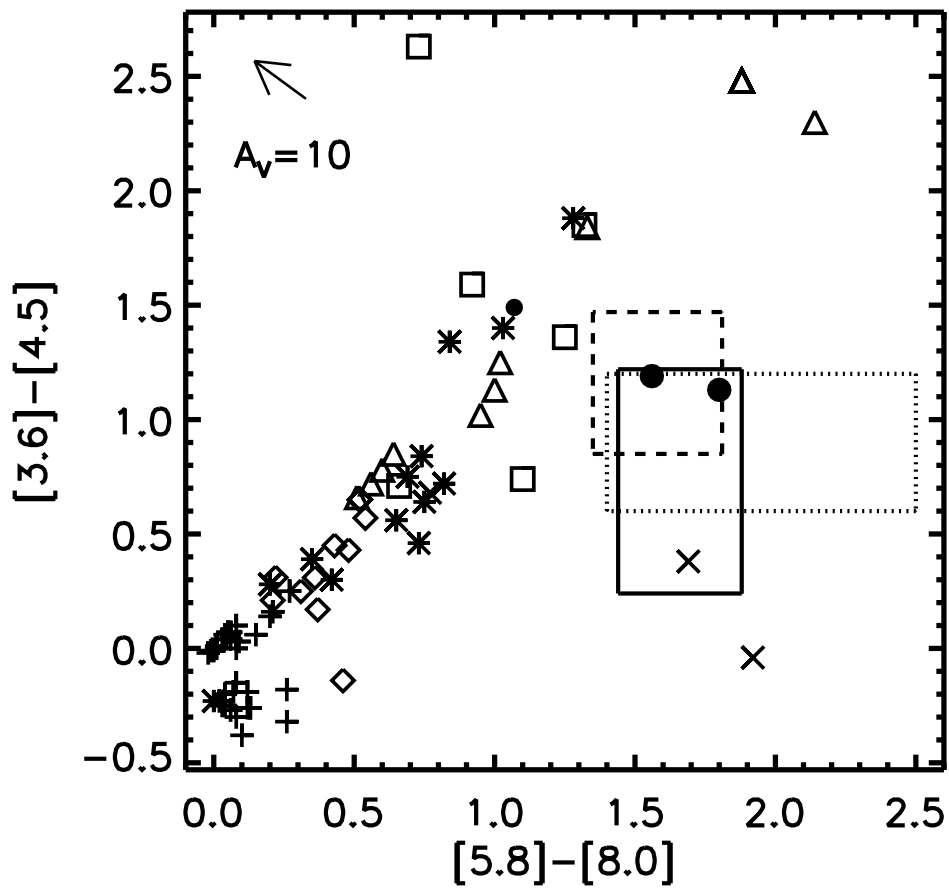} %pn12_341semsws.eps}
\caption{Diagnostic IRAC color-color plot ([3.6]-[4.5] vs. [5.8]-[8.0])
comprising 87 types of IR point source. Key:
pluses - normal dwarfs, giants, supergiants; asterisks - AGB M stars; diamonds - AGB 
visible C stars; triangles - AGB deeply embedded IR C stars; squares - hyperluminous 
objects (these objects include deeply embedded OH/IR stars and early-type hypergiants (Cohen 1993);
a small number are required to reproduce MIR source counts at low latitude: Wainscoat et al. 
1992); crosses - exotica (T Tau stars, reflection nebulae); larger filled circles - 
planetary nebulae; small filled circle - bright compact H{\sc ii} regions.
The reddening vector corresponding to an A$_V$ of 10 mag is shown by the shaft 
of the arrow in the upper left corner.  The three rectangles are described
in the text.  The solid one is that
occupied by the MASH PNe in this paper ($\pm$2 standard errors of the median). 
The July 2004 relative spectral response files were used to represent the IRAC bands.
\label{colcol}}
\end{figure}

Fig.~\ref{colcol} illustrates the ([3.6]-[4.5],[5.8]-[8.0]) color-color 
plane with 87 types of sources plotted.  This diagram too was synthesized from the 
spectral library embedded in the ``SKY" model (Cohen 1993).  Overlaid on
this plane are three boxes for PNe.  The solid box is for the median$\pm$2\,sem
for the sample of MASH PNe.  The dashed box
is for the ISO/SWS sample of PNe but is based on median$\pm$1\,sem colors due to the
poor signal-to-noise ratios at short wavelengths. The third (dotted) box
corresponds to the entire range of colors for the small sample 
of PNe from Hora et al. (2004: their Fig.~3).  There is good overlap between the
three boxes of Galactic PNe.  The intersection of these color-color regions includes
the two large filled circles, that represent SKY's predictions for ensemble averaged colors for ``blue"  and
``red" PNe.  Cohen (1993) lists the nebulae used to produce these PN spectra and Walker et al.
(1989) explain how they are distinguished from one another in $IRAS$ color-color planes.

Can we distinguish PNe from the plethora of other MIR sources using this color-color plane? 
Only a single category among the 87 compact MIR sources in SKY is expected to contaminate any of
the three color zones observed for planetaries, shown by the cross near (1.7,0.4) in Fig.~\ref{colcol}.  
That category corresponds to one type of reflection nebula. This contaminant is easily removed
by comparing H$\alpha$ and red continuum exposures.  Compact H{\sc ii} regions (small filled
circle in the figure) cannot be confused with PNe but some extended H{\sc ii} regions do
overlap the lower right portion of the solid box for MASH PNe (see Cohen et al. 2007; their Fig.~8).
We attribute this similarity of colors to PAHs that dominate the MIR spectra of many H{\sc ii} regions 
and of some PNe.  However, these two classes can be readily separated by MIR morphology and/or by
optical spectroscopy.  Spiral galaxies (both normal and interacting), ellipticals, and irregulars all 
show minimal overlap with MASH PNe and none with the other PN samples, while QSOs do not overlap
any PN color-color zone. (Smith et al. 2007).  

Young stellar objects (YSOs) span a very wide range of GLIMPSE colors.  From a
large grid of pre-computed models (Robitaille et al. 2006) one finds that  
some younger YSOs would overlap with PN colors.  The energy distributions of 
bona fide PNe are never even quasi-continua.  This can lead to real ambiguity when 
model energy distributions are compared with solely broadband measurements of spectra 
that contain strong emission lines, and broad emission or absorption bands.
Consider a confirmed PN whose MIR spectral energy distribution is well-matched in the 
IRAC range by that of a YSO from this grid (Robitaille et al. 2007).  Currently, a true PN
must be optically detected, suffer no unusual reddening,
be isolated from obvious star-forming regions, emit the appropriate optical 
forbidden lines with the correct intensities, and have a morphology plausible for a PN.
Thus, no MASH PN is actually a YSO.  Conversely, one could discriminate between a
true PN and a YSO, whose IRAC photometry places it within the color-color domain that 
we have identified for MASH PNe, by the application of these additional criteria.
Perhaps the simplest test is to examine the extinction required to match a YSO model 
to the IRAC photometry.  Good fits of YSO models to PN energy distributions
often involve high reddening (A$_V$$\geq$5) and the existence of an optical counterpart
makes it less likely that one has found such a YSO.
Colors alone cannot uniquely identify new IR-discovered PNe.  The goal is simply
to isolate plausible candidates for future follow-on spectroscopy.  

\section{MIR/radio flux density ratios}
Radio flux densities were obtained from the MGPS2 (Green 2002, Murphy et al. 2007) 
and NVSS (Condon et al. 1998) surveys. MGPS2 covers the sky south of $-$30$^\circ$ Dec. at 843\,MHz and NVSS covers
the sky north of $-$40$^\circ$ Dec. at 1.4\,GHz.

For MGPS2, flux densities were calculated using the using the {\sc miriad} {\sc imfit}
task to fit elliptical Gaussians above a planar background.
For the NVSS the source catalog was queried and {\sc imfit} was also run on
the survey images as confirmation. The images were inspected to reject
multiple source detections and unrelated chance alignments.  Mauch et al. (2003) 
and Murphy et al. (2007) determined that the flux density 
calibration of NVSS and MGPS2 agreed to within 2\%, based on a 
comparison of $\sim$7000 sources in the overlap zone between these two surveys.

All MGPS2 detections of our PN sample were unresolved (as defined in
Mauch et al. 2003) and hence the fitted peak flux density is used as a 
good measure of the integrated radio fluxes.  NVSS detections were treated 
identically.

All the PNe should have thermal 
spectra in the radio continuum and the difference between flux densities of an optically thin PN
at these two frequencies is about 5\% (after allowance for the Gaunt
factor). If the PN are optically thin, then this change is
smaller than the uncertainties in the radio flux determinations. If the
nebulae were fully optically thick in the radio domain
($S\propto\nu^{\alpha}$), then the flux ratios between 1.4 GHz and 843  MHz
would be 2.8. Taking even this most extreme case, the range of values
for either one of the frequencies is more than any global change in the ratios due
to any mild optical depth effects. Furthermore, the significant difference
between the present PN sample and the 
previous estimate using bright PNe from the Acker et al. (1992) database
remains. For the present analysis, we do not distinguish between results
from these two surveys. The reason for the small number of radio detections  
found partly reflects the confusion from high source density along the
Galactic Plane, but is also due to the intrinsic radio weakness of the MASH PNe.

Might the MASH PNe be optically thick around 1\,GHz?  
Forty-five objects in the compilation of radio measurements of 557 PNe by Higgs (1971)
have either detections or useful limits below 1\,GHz.  We examined the uniform set of models 
for which Higgs adopted T$_e$=12000\,K and fitted for $\tau$(10\,GHz).  The median value
of $\tau$(10\,GHz) for these 45 PNe is 0.0017.  Free-free radio optical depth 
is $\propto$ $\nu^{-2.1}$ so $\tau$(1.4\,GHz) would be $\approx$0.10, and $\tau$(843\,MHz)
$\approx$0.30.  If MASH PNe were similar to these 45 nebulae, then their NVSS flux
densities would be converted to those at 843\,MHz by multiplying by $e^{-0.2}$ or 0.82.
But many MASH PNe are more evolved, fainter, and probably of lower density than 
previously studied objects.  This too lowers their radio optical depths ($\propto$N$_e^2$).

Ten PNe detected by MSX at 8.3\,$\mu$m also have radio fluxes.  Fifteen detected by
IRAC at 8.0\,$\mu$m also have radio fluxes.  We have examined the relationships between these two
pairs of flux densities.  The regression lines for MSX and radio fluxes have a roughly
linear proportionality with an unweighted mean of 6$\pm$1.4 and a median ratio of 5$\pm$1.5.  The
corresponding regression between IRAC and radio fluxes has a mean of 9$\pm$2 and a median of 
5$\pm$2.5.  Given the factor of 0.74 to be applied to IRAC at 8.0\,$\mu$m to correct its diffuse
calibration, the IRAC mean and median would become 6.5$\pm$1.5 and 4$\pm$2, in
the validated MSX calibration basis.  Within their uncertainties, the MSX/radio and IRAC/radio
ratios agree at the 1$\sigma$ level.  Combining them with inverse-variance weighting yields an
overall MIR/radio ratio of 4.6$\pm$1.2 for the MASH PNe.

Cohen \& Green (2001) took a set of 21 PNe from the Acker et al. (1992) catalog and derived a median
ratio of MSX/radio of 12. This is half the median value of 25$\pm$5 they
obtained for H{\sc ii} regions, which has been confirmed by more detailed
analysis (Cohen et al. 2007). The values for the MASH PNe are more than a factor of 2 smaller
than these selected known PNe from Acker et al.  To find an explanation, firstly the MSX/radio flux ratio 
and PN diameters for the 21 previously known PNe, together with our 10 MASH nebulae, were compared.
No trends are seen and the plotted points for the MASH PNe overlap the
distribution for the sample taken from Acker et al. (1992).  Then we examined the two
populations to see whether the samples were comparable in angular size.   
The 21 PNe have a median diameter of 12$\pm$5$^{\prime\prime}$ while the 10 MASH PNe for which we have MSX
and radio detections have a median of 20$\pm$3$^{\prime\prime}$.  
As reported by Parker et al. (2006), the typical MASH PN is significantly larger (and hence generally more evolved)
than those previously listed in the previously known PN compilations (the average diameter of all 905 MASH PNe is
51$^{\prime\prime}$ compared with $<$10$^{\prime\prime}$ for the $\sim$1500 previously known PNe in Acker et al. (1992)).

If the PNe in the two samples are considered to remain with the same
ionized fraction, but the Acker et al. nebulae were to increase their
diameters by 1.67 to match those of the MASH objects, then the mass of
ionized gas in   
those expanded nebulae would be unchanged although the electron densities would fall substantially.
The beam size of the Molonglo Observatory Synthesis Telescope would still exceed the median 
diameter of the MASH PNe so the observed radio fluxes would be unchanged.  However, the 
increase in diameter would diminish the far-UV radiation required to sustain the PDRs, roughly
as the square of the expansion factor (see Fig.~3 of Bernard-Salas \& Tielens (2005)), or a
factor of 2.8. Therefore, one might expect the median MIR/radio ratio to fall by this factor as
fewer UV photons would be available to pump the PAHs into MIR fluorescence.  This would imply
a median ratio for the typically larger MASH PNe of 12/2.8 by comparison with the smaller
Acker et al. (1992) objects, or a ratio of $\sim$4.3. This is almost exactly what we observe.  Consequently,
unlike H{\sc ii} regions, in which there is relatively little evolution in
MIR/radio flux  once they are past the ultra-compact phase, PN evolution is marked by a 
progressively diminishing MIR/radio ratio as nebulae expand, their stars cool, and their PDRs
dwindle.  It is not possible to probe this evolution at the level of individual
PNe using the present data, but the ensemble averages provide adequate
evidence of this phenomenon.

\section{Optical and MIR PN morphologies}
For 17 of the 58 PNe there is simply no hint of either a MIR central star nor of diffuse MIR emission 
associated with the H$\alpha$ object.  The remaining 41 PNe (71\%) are extended IR objects with
a MIR morphology that either matches or complements that in
H$\alpha$.  The most frequent type is that of a PDR in which 8.0-$\mu$m emission occurs 
around parts of the nebular rim but is displaced to the outside of the ionized gas. 
In 11 of the 41 PNe two or three of the shorter IRAC bands appear to follow the 
distribution of H$\alpha$ emission across the PN.  In 5 of the 41 nebulae with MIR extension, 
all four IRAC bands trace the ionized gas.  We have noted the possible presence of a 
candidate central star in the IRAC images for 11 of the 41 PNe.

\section{Conclusions}
Forty-one PNe of our sample of 58 MASH PNe observed by GLIMPSE have MIR counterparts.  For many of
these, it is not possible to extract meaningful quantitative spatially-integrated fluxes
because of the complex structured background in the MIR.

We have shown that optically confirmed PNe exhibit IRAC colors that distinguish them from
other astronomical sources in the [3.6]-[4.5] vs. [5.8]-[8.0] plane.

The ratio of integrated diffuse 8-$\mu$m and radio fluxes is a discriminator between thermal 
and nonthermal emission regions (Cohen \& Green 2001). A ratio of
MIR/radio fluxes of about 25 implies thermal emission, while very small values around 0.06 indicate 
nonthermal processes (Cohen et al. 2007).  PNe are thermal emitters and we have compared the MASH 
sample with the set of bright PNe for which a ratio of 12 was calculated (Cohen \& Green 2001).
The difference in evolutionary state between a typical MASH PN and a nebula drawn from old catalogs
causes the median MIR/radio ratio for our sample of MASH PNe to be about 5.  Evolved PNe are of
lower density, lower UV optical depth, and have only weak or negligible PDRs, reducing
the contribution of PAH emission in bands near 8\,$\mu$m, and hence their MIR detectability with IRAC.
The recognition of this evolutionary trend in MIR properties emphasizes the importance of the 
MASH catalog of newly-discovered PNe.

The fraction of bipolar nebulae within our subset of 58 MASH PNe is 28\%, more than twice
the fraction found for the entire MASH catalog.  This reflects the low scale heights and
short lives of the higher-mass stars that are the progenitors of Type~I PNe.  These are 
preferentially observed by the GLIMPSE survey due to its restriction to Galactic latitudes
of only $\pm$1$^\circ$.  The MIR emission in Type~I nebulae is likely to come from the warm dust 
in the associated circumstellar disks.

False-color IRAC imagery (encoding bands at 4.5, 5.8, 8.0\,$\mu$m as blue, 
green, and red, respectively)
reveals MIR counterparts of PNe that are often clearly distinct from the color of
their surroundings.  The three types of false color found in our survey appear to differentiate
between the degree of excitation of the optical nebular spectra as determined by
[N{\sc ii}]/H$\alpha$.  The most frequently encountered false color
is violet, which represents the weakest optical excitation.  This probably corresponds
most frequently to PNe that contain only H recombination lines.
By combining false color MIR information with the relationship between MIR and H$\alpha$
morphologies of PNe (i.e. whether MIR emission in IRAC bands mimics or is found
outside the H$\alpha$ distribution), one could assess the excitation
of the nebula and decide whether fine-structure
lines produce the emission in the IRAC bands.  For example, the [Ar{\sc ii}] and
[Ar{\sc iii}] lines at 6.99 and 8.99\,$\mu$m, respectively, would contribute to IRAC's
8.0-$\mu$m band as would the 7.64-$\mu$m [Ne{\sc VI}] in very high-excitation PNe.

The ratio of IRAC and MSX flux densities in PNe confirms that a multiplicative correction 
factor of $\sim$0.74 should be applied to IRAC 8.0-$\mu$m diffuse emission fluxes 
to match the absolutely validated calibration of MSX.  These MASH PNe probe the diffuse
calibration on a spatial scale up to 1.3\arcmin, complementing our work on H{\sc ii} 
regions that explored IRAC calibration up to a scale of 24\arcmin\, (Cohen et al. 2007).
This factor is already recommended by the {\it Spitzer} Science Center, and has been derived 
from knowledge of the instrument and from elliptical galaxies.  We independently confirm 
this value for the 8.0-$\mu$m correction factor on a spatial scale between slightly resolved
PNe and well-resolved H{\sc ii} regions.

\section{Acknowledgments}
MC thanks NASA for supporting this work under ADP grant NNG04GD43G with
UC Berkeley.  MC is also grateful for support from the School of Physics in the 
University of Sydney through the Denison Visitor program, and from the Distinguished
Visitor program at the Australia Telescope National Facility in Marsfield.
We thank the referee, Eric Lagadec, for his careful reading of our manuscript.
The MOST is owned and operated by the University of Sydney, with support from the
Australian Research Council and Science Foundation within the School of Physics.
Support for this work, part of the {\it Spitzer} Space Telescope Legacy Science 
Program, was provided by NASA through contracts 1224653 (Univ. of Wisconsin, Madison), 
1224988 (Space Science Institute), 1259516 (UC Berkeley), with the Jet Propulsion Laboratory, 
California Institute of Technology under NASA contract 1407. 
RI acknowledges his funding as a Spitzer Fellow.
This work made use of data products from the Midcourse Space
eXperiment.  Processing of the data was funded by the Ballistic 
Missile Defense Organization with additional support from NASA's Office of Space Science.  
This research has also made use of the NASA/IPAC Infrared Science Archive, which is 
operated by the Jet Propulsion Laboratory, California Institute of Technology, 
under contract with the National Aeronautics and Space Administration.
This research made use of Montage, funded by the
National Aeronautics and Space Administration's Earth Science Technology
Office, Computational Technnologies Project, under Cooperative Agreement
Number NCC5-626 between NASA and the California Institute of Technology.
This research made use of SAOImage {\sc ds9}, developed by Smithsonian
Astrophysical Observatory.

\clearpage

\clearpage

\clearpage

\clearpage

\clearpage

\clearpage

\clearpage

%\clearpage

\clearpage

\clearpage

\clearpage

\clearpage

\clearpage

\clearpage

\clearpage

\clearpage

\begin{deluxetable}{lcrrcccccccccc} 
\tablecolumns{13} 
\tablewidth{0pc}
\setlength{\tabcolsep}{0.025in}
\tabletypesize{\footnotesize} %2column 
\tablecaption{Attributes of PNe studied in H$\alpha$, MIR, and radio. 
Listed are: PN name (col.1); PN status as T(rue), L(ikely), or P(ossible) PN (col.2); coordinates (cols.3-6); apparent size
(col.7); optical morphology type (col.8); whether a PN can be distinguished by false color from its environs (col.9); 
whether optical and IR morphologies are related (col.10); the existence of an MSX 8.3-$\mu$m detection (col.11)
or a radio counterpart (cols.12-13); and whether we have identified an optical or IR central star.  \label{pnelist}}
\tablehead{Name& PN& GLON& GLAT&  RAJ2000& DecJ2000&  Size& Morph.& SST& SST&      MSX& MGPS& NVSS& Star\\
PHR& status&     &     &  &  &          arcsec& type  & col& H$\alpha$&    &     &     &\\}
\startdata

 1806$-$1956&  T&  10.2111&   0.3433&   18h06m55.3s&  $-$19d56m18s& 61$\times$50& Bams&   n&   n&           n&   ...&   n& ...\\
 1807$-$1827&  P&  11.5293&   1.0039&   18h07m11.7s&  $-$18d27m54s& 7$\times$6&    E&     n&   y&           y&   ...&   n& ...\\
 1813$-$1543&  T&  14.6575&   1.0115&   18h13m29.0s&  $-$15d43m19s& 27$\times$21& Eas&    n&   n&           n&   ...&   n&...\\
 1815$-$1457&  P&  15.5185&   1.0342&   18h15m06.5s&  $-$14d57m21s& 9$\times$8& Es&      n/a&  y&           n&   ...&   y&...\\
 1818$-$1526&  L&  15.5378&  $-$0.0195& 18h18m59.2s&  $-$15d26m22s& 55$\times$11& Br?&   n&   y&           n&   ...&    n&..m?\\
 1824$-$1505&  T&  16.4158&  $-$0.9312& 18h24m02.1s&  $-$15d05m33s& 30$\times$18& Bps&   n&   n&           n&   ...&    n&...\\
 1821$-$1353&  P&  17.2190&   0.1272&   18h21m43.9s&  $-$13d53m13s& 20$\times$6& As&     n&   n&           n&   ...&    n&...\\
 1826$-$0953&  T&  21.2911&   0.9803&   18h26m26.1s&  $-$09d53m26s& 54$\times$42&  Bs&   n&   n&           n&   ...&    y&...\\
 1831$-$0805&  L&  23.4401&   0.7449&   18h31m19.6s&  $-$08d05m43s& 13$\times$9& Eas&    y&   y&           n&   ...&    n& N\\
 1834$-$0824&  T&  23.5513&  $-$0.1362& 18h34m41.6s&  $-$08d24m20s& 31$\times$26 & Ea&   n&  y&           n&   ...&    n&...\\
 1842$-$0539&  L&  26.8632&  $-$0.5529& 18h42m17.9s&  $-$05d39m13s& 90$\times$65& Ias&   n&   y&           n&   ...&    n&...\\
 1843$-$0541&  T&  26.9222&  $-$0.7630& 18h43m10.4s&  $-$05d41m51s& 48$\times$39& B?/Eas& n&   y&         n&   ...&    n&...\\
 1844$-$0517&  P&  27.4764&  $-$0.9616& 18h44m54.0s&  $-$05d17m36s& 31$\times$26& Es&    n&   n&          n&   ...&    n&...\\
 1838$-$0417&  P&  27.5860&   1.0186&   18h38m02.2s&  $-$04d17m24s& 15$\times$13& Em&    y&   y&           n&   ...&    n&...\\
 1844$-$0503&  T&  27.6643&  $-$0.8265& 18h44m45.7s&  $-$05d03m54s& 35$\times$12& Bm?/Em& y&   y&           y&   ...&   n& oi\\ 
 1844$-$0452&  L&  27.7721&  $-$0.6350& 18h44m17.3s&  $-$04d52m56s& 38$\times$38& Rar&    n&  n&           n&   ...&    n& m\\
 1845$-$0343&  L&  28.8931&  $-$0.2907& 18h45m06.0s&  $-$03d43m33s& 51$\times$30& As&    y&   y&           n&   ...&    n& oi?\\
 1843$-$0325&  P&  28.9519&   0.2570&   18h43m15.3s&  $-$03d25m27s& 10$\times$9& Ea&     y&   y&           y&   ...&    y?& N\\
 1842$-$0246&  L&  29.5024&   0.6246&   18h42m57.1s&  $-$02d46m01s& 24$\times$13& Em&    y&   y&           n&   ...&    n&...\\
 1843$-$0232&  T&  29.8197&   0.5073&   18h43m56.9s&  $-$02d32m08s& 61$\times$54& Ear&   n&   y&           n&   ...&    n&...\\ 
 1846$-$0233&  T&  30.0485&   0.0357&   18h46m02.7s&  $-$02d33m09s& 36$\times$31& Ear&   n&   y&           n&   ...& n\\
 1847$-$0215&  T&  30.5060&  $-$0.2200& 18h47m47.4s&  $-$02d15m30s& 20$\times$14& Bs&    y&   n&           n&   ...&    n& m?\\
 1856+0028&  L&  33.9770&  $-$0.9860&   18h56m51.1s&    +00d28m53s& 8$\times$7&  Er&     y&   y&           y&   ...&    n&...\\
 1857+0207&  L&  35.5650&  $-$0.4910&   18h57m59.5s&    +02d07m07s& 11$\times$11&  Ea&   y&   y&           y&   ...&    y& N\\
 1150$-$6226& L&  295.9050&  $-$0.4110& 11h50m07.0s&  $-$62d26m32s& 89$\times$77&  R&    n&   n&           n&   n&    ...&...\\
 1152$-$6234& L&  296.2510&  $-$0.4580& 11h52m55.7s&  $-$62d34m10s& 27$\times$24&  R&    n&   n&           n&   n&    ...&...\\
 1157$-$6312& L&  296.8490&  $-$0.9840& 11h57m03.2s&  $-$63d12m44s& 15$\times$13&  B?&   y&   y&           y&      y&    ...&...\\
 1206$-$6122& T&  297.5680&   1.0230&   12h06m25.5s&  $-$61d22m44s& 2$\times$11&  E&     y&   y&           n&   n&    ...&...\\
 1218$-$6245& L&  299.1190&  $-$0.1360& 12h18m00.9s&  $-$62d45m38s& 42$\times$31&  Es&   n&   y&           n&   ...&    ...&...\\
 1223$-$6236& T&  299.7780&   0.0980&   12h23m58.0s&  $-$62d36m21s& 48$\times$43&  E&    n&   y&            n&   n&    ...&...\\
 1244$-$6231& T&  302.1330&   0.3510&   12h44m28.5s&  $-$62d31m19s& 300$\times$235&  B&  n&   y&        n&   n&    ...& o$^b$\\ 
 1246$-$6324& T&  302.3730&  $-$0.5390& 12h46m26.5s&  $-$63d24m28s& 31$\times$19&  B&    y&   y&        y&   y&    ...& oim\\
 1250$-$6346& T&  302.7840&  $-$0.9080& 12h50m04.4s&  $-$63d46m52s& 83$\times$74&  Ea&   n&   y&        n&   n&    ...& o$^b$?m\\
 1255$-$6251& L&  303.3725&   0.0173&   12h55m18.0s&  $-$62d51m04s& 185$\times$81&  B&   n&   y&        n&   n&    ...& oi\\
 1257$-$6216& P&  303.6783&   0.5923&   12h57m51.3s&  $-$62d16m12s& 19$\times$13&  E&    y&   y&        n&   n&    ...& m\\
 1408$-$6229& T&  311.7300&  $-$0.9500& 14h08m47.3s&  $-$62d29m58s& 82$\times$46&  B&    n&   y&        n&   n&    ...& m\\
 1408$-$6106& T&  312.1525&   0.3741&   14h08m51.7s&  $-$61d06m27s& 307$\times$264& Es&  n&   y&       n&   n&    ...& m?\\
 1429$-$6043& P&  314.6780&  $-$0.1290& 14h29m52.8s&  $-$60d43m57s& 167$\times$131&   E& n&   n&        n&   n&    ...& m\\
 1429$-$6003& L&  314.9220&   0.5180&   14h29m43.9s&  $-$60d03m17s& 141$\times$110&  Ems& n&   y&        n&   n&    ...&...\\
 1437$-$5949& T&  315.9480&   0.3320&   14h37m53.2s&  $-$59d49m25s& 103$\times$63&  Ba&  n&   y&        n&   y&    ...&...\\
 1447$-$5838& P&  317.5785&   0.8845&   14h47m41.8s&  $-$58d38m41s& 200$\times$200&  I/Ba&    n&   n&        n&   n& ...& o$^b$i\\ 
 1457$-$5812& T&  318.9300&   0.6930&   14h57m35.8s&  $-$58d12m09s& 31$\times$25&  A&   y&   y&        y&   y&    ...& N\\
 1507$-$5925& T&  319.5050&  $-$1.0140& 15h07m50.2s&  $-$59d25m14s& 22$\times$17&  Ea&   y&   y&        y&   y&    ...& o\\
 1544$-$5607& P&  325.4480&  $-$1.0270& 15h44m56.7s&  $-$56d07m07s& 14$\times$10&  E&    n&   n&        n&   n&    ...&...\\
 1552$-$5254& T&  328.3570&   0.7670&   15h52m56.8s&  $-$52d54m12s& 31$\times$26&  Es&   y&   y&       y&   y&    ...&...\\
 1610$-$5130& P&  331.2780&   0.0600&   16h10m21.1s&  $-$51d30m54s& 20$\times$11&  Es&   n&   n&       n&   n&    ...& m\\
 1619$-$5131& P&  332.3493&  $-$0.9814& 16h19m57.6s&  $-$51d31m48s& 11$\times$11&  E&    y&   y&        n&   y&    ...&...\\
 1622$-$5038& L&  333.2746&  $-$0.6547& 16h22m40.6s&  $-$50d38m42s& 21$\times$19&  Ear&  y&   y&        y&   n&    ...&...\\
 1619$-$4914& T&  333.9279&   0.6858&   16h19m40.1s&  $-$49d14m00s& 36$\times$32&  Rs&   y&   y&        y&   y&    ...& oiNm\\ 
 1619$-$4907& T&  334.0350&   0.7560&   16h19m50.1s&  $-$49d06m52s& 48$\times$47&  Ra&    n&   y&        n&   n&    ...&...\\
 1633$-$4650& T&  337.3141&   0.6361&   16h33m58.0s&  $-$46d50m07s& 24$\times$8&  B&      y&   y&        y&   y&    ...&...\\
 1635$-$4654& P&  337.4831&   0.3524&   16h35m51.9s&  $-$46d54m10s& 77$\times$34& Ia/B?&  n&   n&        n&   n&    ...& o$^b$\\
 1634$-$4628& T&  337.6825&   0.7684&   16h34m51.2s&  $-$46d28m28s& 22$\times$17&  E&     n&   y&        n&   n&    ...& m\\
 1639$-$4516& T&  339.0980&   0.9880&   16h39m22.3s&  $-$45d16m35s& 38$\times$25&  Er&    n&   n&        n&   n&    ...&...\\
 1644$-$4455& P&  339.9730&   0.5280&   16h44m36.2s&  $-$44d55m23s& 34$\times$34&  E&     n&   y&        y&   n&    ...& m\\
 1646$-$4402& L&  340.8600&   0.8500&   16h46m27.6s&  $-$44d02m25s& 71$\times$72&  A&     n&   y&        n&   n&    ...&...\\
 1709$-$3931& T&  347.0320&   0.3500&   17h09m10.8s&  $-$39d31m06s& 51$\times$13&  B&     n&   n&        n&   n&    ...& oi\\
 1714$-$4006& T&  347.2000&  $-$0.8720& 17h14m49.3s&  $-$40d06m09s& 20$\times$11&  B?&    y&   y&        n&   y&    ...& m\\
\enddata 
\tablenotetext{b}{CS candidate is blue.}
%\end{center} 
\end{deluxetable}

\clearpage

\begin{deluxetable}{lccccrrrrrr} 
\tablecolumns{10} 
\tablewidth{0pc}
\tabletypesize{\scriptsize} 
\tablecaption{Radio and MIR flux densities for PNe with two or more detections of these. 
Listed are: PN name (col.1); PN status (col.2); representative diameter (col.3); optical morphology
(col.4); radio detection by M(GPS2) or N(VSS) (col.5); radio flux density with parenthesized
uncertainty (col.6); MSX and IRAC flux densities (cols. 7-8); ratio of IRAC to MSX flux (col.9); 
ratio of MSX to radio flux (col.10); ratio of IRAC to radio flux (col.11). \label{fluxes}} 
\tablehead{PN NAME&  Status&  Diam.&   Morph.& [M]GPS2&  Radio&  MSX8.3&  IRAC8.0& IRAC/& MSX/& IRAC/\\
           PHR&           & arcsec&    type& [N]VSS& mJy&    mJy&    mJy&  MSX& radio & radio\\}
\startdata
 1844$-$0503& T&  20&    Bm?/Em& N& 3.2(0.9)&      43&         78& 1.8& 13.0& 24.0\\ 
 1843$-$0325& P&  9&	  Ea&   N&  16.0(2.0)&    210&        210& 1.0& 13.0&  13.0\\
 1857+0207& L&  11&    Ea&   N&  78.3(2.4)&    160&        180& 1.1& 2.0& 2.3\\
 1246$-$6324& T&  24&    B&  M&    14.3(1.4)&    120&        140&  1.2&  8.2& 10.0\\
 1457$-$5812& T&  28&    A&  M&    79.8(2.7)&       78&        160& 2.1& 1.0& 2.0\\
 1507$-$5925& T&  19&    Ea& M&  24.2(1.3)&      110&        100& 0.9& 5.4& 4.8\\
 1552$-$5254& T&  28&    Es& M&   20.9(2.2)&      110&        150& 1.4& 4.2& 5.9\\
 1619$-$4914& T&  34&    Rs& M& 240.(8.0)&      1000&        970& 1.0& 4.3& 4.1\\
 1619$-$5131& P&  11&    E& M&  20.2(1.8)&       90&         99& 1.1& 4.5& 4.9\\
 1633$-$4650& T&  14&    B& M&   26.9(3.1)&       55&         63& 1.2& 2.0& 2.3\\
 1714$-$4006& T&   15&  B?& M& 9.7(1.3)&       $<$31&  27& \nodata& \nodata&   2.8\\ 
 1157$-$6312& L&  13&    B?&  M&    6.4(1.0)&   $<$20&  30& \nodata& \nodata& 4.7\\
 1437$-$5949& T&  181&   Ba&  M&  11.7(1.8)&     $<$160&   320& \nodata& \nodata& 27.0\\
 1815$-$1457& P&  8&	  Es& N&    9.2(0.6)&     $<$26&     16& \nodata& \nodata& 1.7\\
 1826$-$0953& T&  48&    Bs& N&    8.4(0.6)&   $<$160&     170& \nodata& \nodata& 20.0\\
 1644$-$4455& P&  34&     E& M&    $<$5& 150&     150& 1.0& \nodata& \nodata\\
 1843$-$0541& T&  43&   B?/Eas& N& $<$1.5&    130&   180&   1.4& \nodata& \nodata\\
 1622$-$5038& L&   20&   Ear& M&   $<$28&       51&   110& 2.2&  \nodata& \nodata\\
 1842$-$0539& L&  77&    Ias& N&    $<$2.5&      280&   370& 1.3& \nodata& \nodata\\
\enddata
\end{deluxetable} 

\clearpage

\begin{deluxetable}{ccc}
\tablecolumns{3} 
\tablewidth{0pc}
\tablecaption{Observed and corrected median colors (in magnitudes) of MASH PNe in 
the four IRAC bands, compared with colors synthesized from ISO SWS spectra of a
sample of well-known Acker et al. (1992) PNe. \label{colors} }
\tablehead{Color&       MASH&          Acker et al.\\
           Index&    Corrected&        synthesized\\
                &    median$\pm$sem&   median$\pm$sem\\}
\startdata
$[3.6]-[4.5]$&    0.68$\pm$0.22&  1.16$\pm$0.31\\ 
$[3.6]-[5.8]$&    2.12$\pm$0.25&  2.37$\pm$0.20\\    
$[3.6]-[8.0]$&    3.84$\pm$0.28&  3.91$\pm$0.28\\
$[4.5]-[5.8]$&    1.25$\pm$0.16&  1.07$\pm$0.28\\
$[4.5]-[8.0]$&    2.76$\pm$0.19&  2.57$\pm$0.34\\
$[5.8]-[8.0]$&    1.66$\pm$0.11&  1.62$\pm$0.23\\
\enddata
\end{deluxetable}

\end{document}